\pdfoutput=1

\documentclass[11pt,twoside,a4paper,cmspaper,final,collab]{cms-tdr}
\def\svnVersion{289038}\def\cmsCernNo{2013-037}\def\cmsCernDate{\today}\def\cmsMessage{Published in the Journal of High Energy Physics as \href{http://dx.doi.org/10.1007/JHEP04(2015)124}{\doi{10.1007/JHEP04(2015)124.}}}

\begin{document}\cmsNoteHeader{SUS-14-014}

\hyphenation{had-ron-i-za-tion}
\hyphenation{cal-or-i-me-ter}
\hyphenation{de-vices}

\RCS$Revision: 282464 $
\RCS$HeadURL: svn+ssh://svn.cern.ch/reps/tdr2/papers/SUS-14-014/trunk/SUS-14-014.tex $
\RCS$Id: SUS-14-014.tex 282464 2015-03-26 19:06:48Z alverson $
\newlength\cmsFigWidth
\ifthenelse{\boolean{cms@external}}{\setlength\cmsFigWidth{0.85\columnwidth}}{\setlength\cmsFigWidth{0.4\textwidth}}
\ifthenelse{\boolean{cms@external}}{\providecommand{\cmsLeft}{top}}{\providecommand{\cmsLeft}{left}}
\ifthenelse{\boolean{cms@external}}{\providecommand{\cmsRight}{bottom}}{\providecommand{\cmsRight}{right}}

\newcommand{\njets}{\ensuremath{N_\text{jets}}}
\newcommand{\Rinout}{\ensuremath{R_\text{out/in}}\xspace}
\newcommand{\Rsfof}{\ensuremath{R_\text{SF/OF}}\xspace}
\newcommand{\Reeof}{\ensuremath{R_\text{\Pe\Pe/OF}}\xspace}
\newcommand{\Rmmof}{\ensuremath{R_\text{\Pgm\Pgm/OF}}\xspace}
\newcommand{\rmue}{\ensuremath{r_{\Pgm{}\Pe}}\xspace}
\newcommand{\RT}{\ensuremath{R_\text{T}}\xspace}
\newcommand{\etalep}{\ensuremath{\eta_\text{lep}}\xspace}
\newcommand{\EM}{\ensuremath{\Pe^\pm\Pgm^\mp}\xspace}
\newcommand{\wjets}{\ensuremath{\PW+\text{jets}}\xspace}
\newcommand{\zjets}{\ensuremath{\text{DY}+\text{jets}}\xspace}
\newcommand{\DYjets}{\ensuremath{\text{DY}+\text{jets}}\xspace}
\newcommand{\gjets}{\ensuremath{\gamma+\text{jets}}\xspace}
\newcommand{\ZZ}{\ensuremath{\Z\Z}\xspace}
\newcommand{\Zg}{\ensuremath{\Z/\gamma^{*}}\xspace}
\newcommand{\lumifinal}{19.4\fbinv\xspace}
\newcommand{\significanceNumber}{$2.6$}
\newcommand{\mll}{\ensuremath{m_{\ell\ell}}\xspace}
\newcommand{\JZB}{\ensuremath{\mathrm{JZB}}\xspace}
\newcommand{\central}{central\xspace}
\newcommand{\forward}{forward\xspace}
\newcommand{\slepton}{\ensuremath{{\widetilde{\ell}}}\xspace}
\newcommand{\secondchi}{\PSGczDt\xspace}
\newcommand{\firstchi}{\PSGczDo\xspace}
\newcommand{\sbottom}{\PSQb\xspace}

\cmsNoteHeader{SUS-14-014}
\title{Search for physics beyond the standard model in events with two leptons, jets, and missing transverse momentum in pp collisions at $\sqrt{s}=8$\TeV}

\date{\today}

\abstract{
A search is presented for physics beyond the standard model in final states with two opposite-sign same-flavor leptons, jets, and missing transverse momentum. The data sample corresponds to an integrated luminosity of 19.4\fbinv of proton-proton collisions at $\sqrt{s}=8$\TeV collected with the CMS detector at the CERN LHC in 2012. The analysis focuses on searches for a kinematic edge in the invariant mass distribution of the opposite-sign same-flavor lepton pair and for final states with an on-shell Z boson. The observations are consistent with expectations from standard model processes and are interpreted in terms of upper limits on the production of supersymmetric particles.
}

\hypersetup{%
pdfauthor={CMS Collaboration},%
pdftitle={Search for physics beyond the standard model in events with two leptons, jets, and missing transverse momentum in pp collisions at sqrt(s) = 8 TeV},%
pdfsubject={CMS},%
pdfkeywords={CMS, physics, supersymmetry, SUSY}}

\maketitle

\section{Introduction}
\label{sec:intro}

This paper presents a search for physics beyond the standard model (SM)
in events containing a pair of opposite-sign same-flavor (SF) electrons or muons, jets, and an imbalance in transverse momentum.
The analysis is based on a sample of proton-proton (pp) collisions collected at a center-of-mass energy of 8\TeV with the
CMS detector~\cite{CMS:2008zzk} at the CERN LHC in 2012 and corresponds to an integrated luminosity of $\lumifinal$.

The invariant mass distribution of the two-lepton system can exhibit an excess that increases with the dilepton mass, followed by a sharp decrease and thus an ``edge'',
if the two leptons originate from the decay of an on-shell heavy neutral particle. This kind of signature
is fairly generic for models of physics beyond the SM (BSM), assumes an isotropic decay, and is
purely kinematic in origin.
In models of supersymmetry (SUSY)~\cite{Martin:1997ns}, an edge with a triangular shape is expected in the cascade process
$\PSGczDt \to \ell\slepton \to \PSGczDo \ell^+\ell^-$~\cite{edge}, where $\PSGczDt$ and $\PSGczDo$ are respectively the
 next-to-lightest and lightest neutralino,
with $\slepton$ a selectron or smuon, the SUSY partners of an electron or muon.
Alternatively, the  $\PSGczDt$ can undergo a three-body decay to $\PSGczDo \ell^+\ell^-$ through a virtual $\Z^{*}$ boson, also yielding an edge in
the dilepton mass spectrum but with a more rounded shape.  Another possibility is the decay of a $\PSGczDt$ to an on-shell Z boson,
$\PSGczDt \to  \PSGczDo \Z$.
This latter process does not produce an edge but rather a dilepton mass peak near 91\GeV.
These processes arise as a consequence of the gauge-coupling structure of SUSY and are a characteristic feature of SUSY decay chains.
Their relative importance depends on the SUSY mass hierarchy and is thus model dependent.

This search is therefore motivated by the possible existence of the fairly generic signal shape of an edge, or of a peak at the Z boson mass,
that would be visible in the invariant mass distribution of the two leptons.
The position of the edge would give an indication of the unknown BSM mass hierarchy.
The dilepton invariant mass provides a search variable that is unaffected by uncertainties in the jet
energy scale and resolution, and the search for an edge is therefore complementary to searches based solely on hadronic quantities.

The CMS Collaboration previously presented two searches for BSM physics based on events with an opposite-sign SF lepton pair:
a search for an edge in the dilepton mass spectrum outside the Z boson mass region~\cite{edge2011},
and a search for events containing on-shell Z bosons~\cite{zjets2011}.
Both these studies were conducted using the 7 TeV CMS data sample collected in 2011.
The present study updates and combines these two analyses, using the 8 TeV data sample. Searches for SUSY in opposite-sign dilepton final states, but targeting different production and/or decay mechanisms, are presented by the ATLAS Collaboration in Refs.~\cite{ATLAS8TeVref1,ATLAS8TeVref2}.

A brief description of the CMS detector is given in Section~\ref{cmsdet}. Signal models studied in this analysis are described in Section~\ref{sec:signal}.
Section~\ref{sec:eventsel} outlines the event selection and simulation.
The background estimation methods are presented in Section~\ref{sec:bkgd}.  A fitting procedure used to search for an edge is described in Section~\ref{sec:kinfit}. The results of the search are presented in Section~\ref{sec:results}.
Systematic uncertainties associated with the signal modeling are discussed in Section~\ref{sec:systematics} and the results of the search are interpreted in the context of the signal models in Section~\ref{sec:interpretation}. A summary is presented in Section~\ref{sec:conclusion}.

\section{Detector and trigger}
\label{cmsdet}
The central feature of the CMS detector is a superconducting solenoid of 6\unit{m} internal diameter that produces an axial magnetic field of 3.8\unit{T}.
A silicon pixel and strip tracker, a lead tungstate crystal electromagnetic calorimeter, and a brass/plastic-scintillator hadron calorimeter
are positioned within the field volume.
Iron and quartz-fiber hadron calorimeters are located outside the magnetic field volume, within each endcap region of the detector.
Muons are measured using gas-ionization detectors embedded in the steel flux-return yoke outside of the solenoid.
A detailed description of the CMS detector, its coordinate system, and the main kinematic variables used in the analysis can be found in Ref.~\cite{CMS:2008zzk}.

Events must satisfy at least one of a set of $\Pe\Pe$, $\Pgm\Pgm$, and $\Pe\Pgm$ triggers.
The $\Pe\Pe$ and $\Pgm\Pgm$ triggers collect signal candidate data while the $\Pe\Pgm$ trigger collects data used in the background-determination procedure, as described below.
These triggers require at least one electron or muon with transverse momentum $\pt>17\GeV$, and another with $\pt>8\GeV$.
Their efficiencies after event selection ($>$90\%) are measured in data and are accounted for in the analysis.
The efficiencies of the $\Pe\Pe$, $\Pgm\Pgm$, and $\Pe\Pgm$ triggers differ by only a few percent in the kinematic range of this search.

\section{Signal scenarios}
\label{sec:signal}
Two classes of signal events are considered, as explained below.
Both classes are implemented in the framework of simplified models~\cite{Alves:2011wf}, in which only the targeted production and decay schemes are examined,
with all non-participating BSM particles assumed to be too heavy to be relevant.

\begin{figure}
\centering
\includegraphics[width=0.49\textwidth]{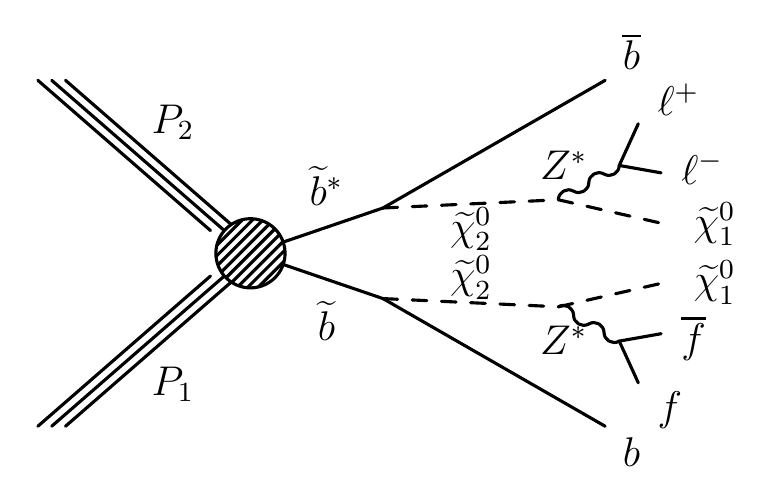}
\includegraphics[width=0.49\textwidth]{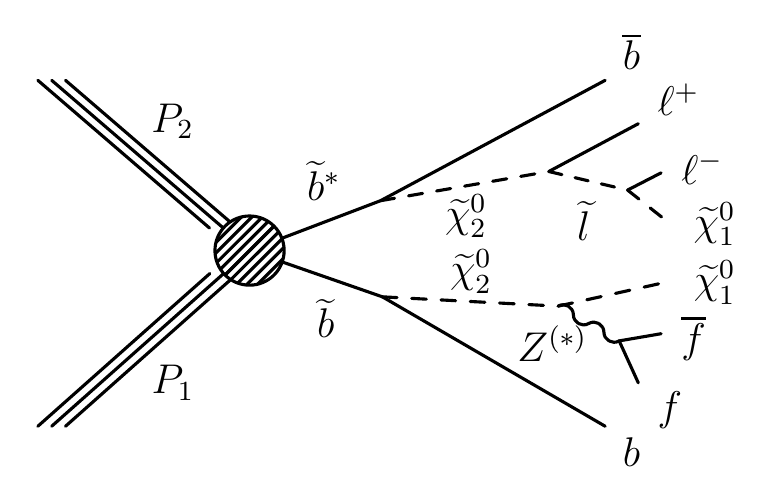}
\caption{
Event diagrams for the (left) ``fixed-edge'', and (right) ``slepton-edge'' scenarios, with
  \sbottom a bottom squark, \secondchi the second lightest neutralino,
  \firstchi a massive neutralino LSP, and $\slepton$ an electron- or muon-type slepton. For the slepton-edge scenario, the
  \Z boson can be either on- or off-shell, while for the fixed-edge scenario it is off-shell.
}
\label{fig:feynmanEdge}
\end{figure}

The first class of signal events targets the production of an edge in the invariant mass spectrum of opposite-sign SF lepton pairs,
as expected from the correlated production of these leptons in cascade decays.
This class of scenarios is based on the production of a bottom squark-antisquark pair.
Each bottom squark \sbottom decays to a bottom quark b and the \secondchi neutralino.
Two specific possibilities are considered.  In the first scenario (Fig.~\ref{fig:feynmanEdge} left), the \secondchi decays to an off-shell \Z boson $\mathrm{Z}^{*}$ and the \firstchi neutralino,
where the \firstchi is a stable, weakly interacting, lightest SUSY particle (LSP). The $\mathrm{Z}^*$ boson decays according to its SM branching fractions, sometimes producing a charged lepton pair $\ell^+\ell^-$ ($\ell = \Pe$, $\mu$). The mass difference between the \secondchi and \firstchi, which determines the location of the edge, is fixed to 70\GeV.
This scenario is referred to as the ``fixed-edge'' scenario.
In the second scenario (Fig.~\ref{fig:feynmanEdge} right), the \secondchi decays to an on- or off-shell \Z boson and the \firstchi LSP or according to $\secondchi \to \ell\slepton$,
with a 50\% probability for each decay. The slepton $\slepton$, \ie, the SUSY partner of the lepton, then decays according to $\slepton \to \ell \firstchi$.
The considered sleptons are mass-degenerate selectrons and smuons.
The mass of the slepton is chosen to lie halfway between the masses of the two neutralinos: $m_{\slepton} = m_{\firstchi} + 0.5(m_{\secondchi}-m_{\firstchi})$.
The mass of the \firstchi is set to 100\GeV, with the position of the edge a free parameter in a scan of the mass spectrum.
This scenario is referred to as the ``slepton-edge'' scenario.

\begin{figure}
\centering
  \includegraphics[width=0.49\textwidth]{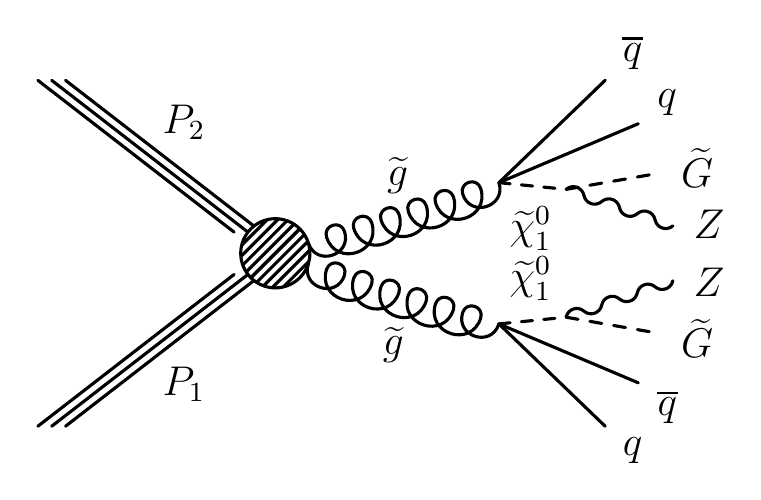}
\caption{
Event diagram for the ``GMSB'' scenario, with $\PSg$ a gluino, \firstchi the lightest neutralino,
   and $\PXXSG$ a massless gravitino LSP.
}
\label{fig:feynmanT5ZZgmsb}
\end{figure}

The second class of signal events targets the production of an  opposite-sign SF lepton pair from the decay of an on-shell \Z boson.
This class of scenarios, illustrated in Fig.~\ref{fig:feynmanT5ZZgmsb}, is based on gluino pair production in the context of
gauge mediated supersymmetry breaking (GMSB) models~\cite{Matchev:1999ft,Meade:2009qv,ref:ewkino}.
Each gluino decays to a quark-antiquark pair and the \firstchi neutralino.
The \firstchi decays to an on-shell \Z boson and a stable, massless, weakly interacting gravitino LSP.
We refer to this scenario as the ``GMSB'' scenario.

The production of squark and gluino pairs is simulated with the \MADGRAPH5.1.3.30~\cite{madgraph5,Alwall:2014hca} Monte Carlo (MC) leading-order event generator,
including up to two additional
partons at the matrix element level. The decays of the squarks, gluinos, and other particles are simulated with the \PYTHIA6.4.22~\cite{Pythia} event generator.
The \MADGRAPH  events are subsequently processed with the \PYTHIA program to generate parton showers and account for hadronization.
The decay of the \Z boson is handled in \PYTHIA.
In all the aforementioned scenarios (fixed-edge, slepton-edge, GMSB), the \Z boson decays according to its SM branching fractions.
To reduce computational requirements, the detector response is simulated using the CMS fast simulation~\cite{CMSFastSim}.
Differences in the lepton reconstruction and identification efficiencies between the fast and a ``full'' simulation,
where the full simulation is based on processing through the \GEANTfour~\cite{Geant} programs, are corrected using scale factors.
The expected signal event rates are normalized to cross sections
calculated at the next-to-leading order (NLO) in the strong coupling constant, including the resummation of soft gluon emission at
next-to-leading-logarithmic (NLO+NLL) accuracy~\cite{bib-nlo-nll-01,bib-nlo-nll-02,bib-nlo-nll-03,bib-nlo-nll-04,bib-nlo-nll-05,ref:xsec}.

\section{Event selection, reconstruction, and search strategy}
\label{sec:eventsel}

We select events with an oppositely charged lepton pair (\EE, \EM, or \MM). The leptons are required to have $\pt>20\GeV$ and $\abs{\eta}<2.4$,
where $\eta$ is the pseudorapidity~\cite{CMS:2008zzk}. In events with more than two selected leptons, we choose the two oppositely charged leptons with highest \pt.
The background estimation techniques employed in this analysis rely, in part, on a symmetry between lepton pairs
with the same flavor compared to those with opposite flavor (OF),
where OF refers to \EM combinations.  It is therefore desirable that the reconstruction efficiencies of electrons~\cite{EGMpaper} and muons~\cite{MUOpaper} be
as similar as possible.  For this reason, we exclude leptons in the intervals $1.4<\abs{\eta}<1.6$ between the barrel and endcap regions of the
detector~\cite{CMS:2008zzk}, where the electron and muon reconstruction efficiencies differ significantly.
For the SF signal candidate sample, only events with an $\EE$ or $\MM$ pair are used.

Leptons produced in the decays of low-mass particles, such as hadrons containing $\cPqb$ and $\cPqc$ quarks, almost always lie in or near jets.
The background from these low-mass processes can be suppressed by requiring the leptons to be isolated in space from other particles.
A cone of radius $\Delta{}R\equiv\sqrt{\smash[b]{(\Delta\eta)^2+(\Delta\phi)^2}}=0.3$ is constructed around the
lepton momentum direction, where $\phi$ is the azimuthal angle.  The lepton relative isolation is then quantified by
the scalar \pt sum of all particle candidates within this cone, excluding the lepton, divided
by the lepton \pt.
The resulting quantity is required to be less than 0.15.
The sum includes a correction to the total energy to account for contributions from extraneous pp interactions within the same or a nearby
bunch crossing (pileup). For electrons, the pileup correction is based on the effective area method~\cite{FastJet}, while for muons it is based on the number
of charged hadrons not associated with the primary vertex.
The performance of the electron and muon isolation criteria is discussed in Refs.~\cite{EGMpaper, MUOpaper}.

The primary vertex is taken to be the reconstructed vertex with the largest $\pt^2$ sum of associated tracks.
Leptons with impact parameters larger than 0.2\mm in the transverse plane or 1\mm along the beam direction are rejected.
As an additional requirement, the two selected leptons must be separated by $\Delta{}R>0.3$ to avoid systematic effects that arise for isolation requirements
in very collinear topologies.

\begin{figure}[!htb]
\begin{center}
  \includegraphics[width=0.49\textwidth]{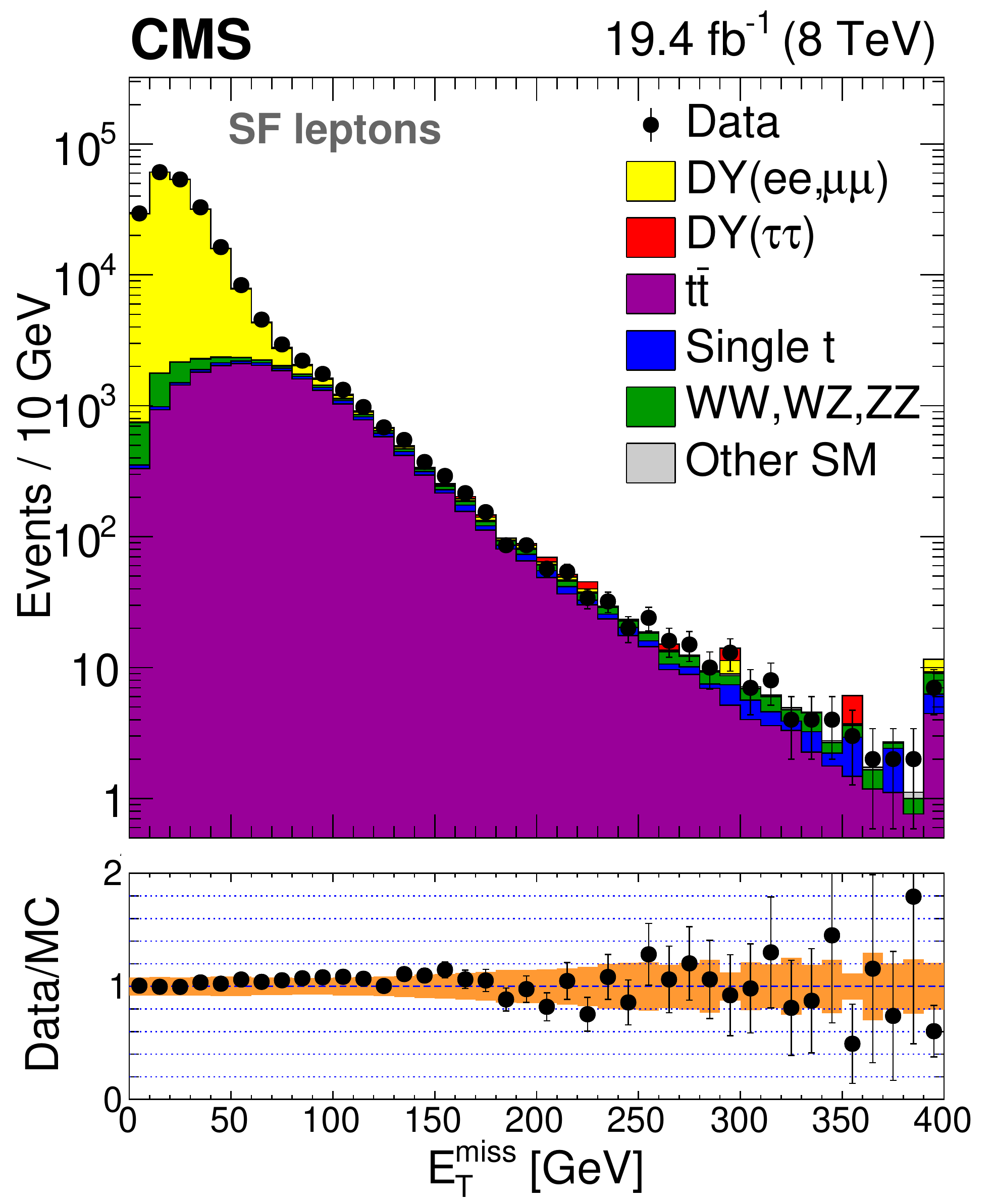}
  \includegraphics[width=0.49\textwidth]{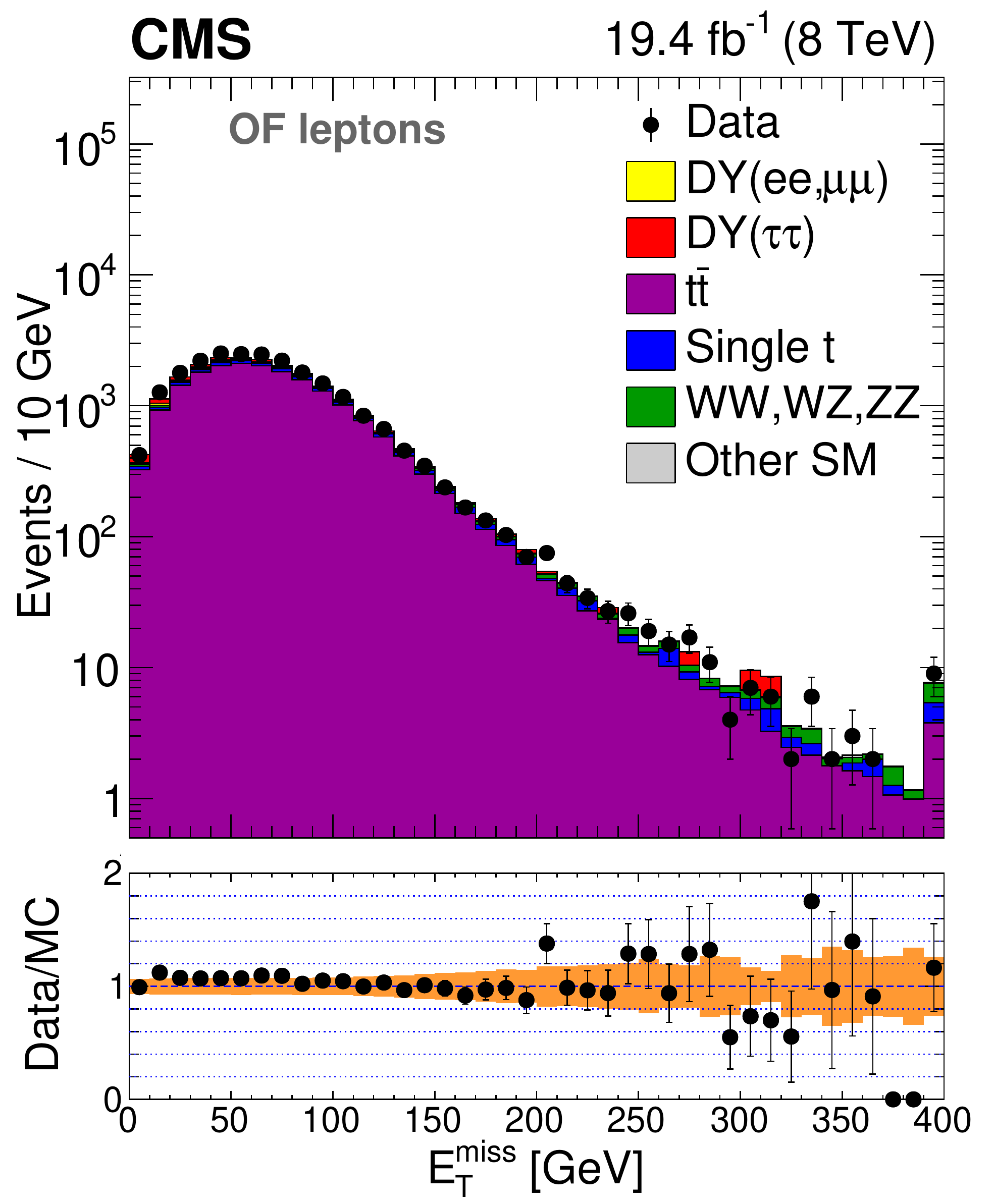}
  \caption{
The \MET distributions of events in the SF (left) and OF (right) samples for $m_{\ell\ell}>20\GeV$ and $N_\text{jets} \geq 2$
in comparison with predictions for the SM background from the MC generators described in Section~\ref{sec:eventsel}.
In the ratio panel below each plot, the error bars on the black points show the statistical uncertainties
of the data and MC samples, while the shaded band indicates the MC statistical and systematic uncertainties added in quadrature.
The rightmost bins contain the overflow.
}
  \label{fig:dataMC_incl}
\end{center}
\end{figure}

A particle-flow (PF) technique~\cite{Chatrchyan:2012zz} is used to reconstruct jets and missing transverse momentum.
All objects reconstructed with the PF algorithm serve as input for jet reconstruction, based on the anti-\kt clustering
algorithm~\cite{Cacciari:2008gp}  with a distance parameter of 0.5, as implemented in the \textsc{FastJet}
package~\cite{Cacciari:2005hq,FastJet}.
We apply \pt- and $\eta$-dependent corrections to account
for residual effects of nonuniform detector response.
The contribution to the jet energy from
pileup is estimated on an event-by-event basis using the jet area method described in
Ref.~\cite{cacciari-2008-659}, and is subtracted from the overall jet \pt.
Jets are required to have  $\pt > 40\GeV$, $\abs{\eta} < 3.0$, and to be separated  by $\Delta{}R>0.4$
from all selected leptons.
The missing transverse momentum
$\ptvecmiss$
is defined as the projection on the plane perpendicular to the beam axis of the negative vector sum of the momenta of all
reconstructed PF objects in an event. The magnitude of
$\ptvecmiss$
is referred to as $\ETmiss$.
The $\ETmiss$ distributions of events in the SF and OF samples for dilepton invariant mass
$m_{\ell\ell}>20\GeV$ and number of jets $N_\text{jets} \geq 2$ are shown in Fig.~\ref{fig:dataMC_incl}.

The event selection criteria are motivated by the expectation that BSM signal events, involving the production of new heavy particles, generally
have larger jet multiplicity and \MET than background events, which primarily arise from top quark-antiquark (\ttbar) and Drell--Yan (DY) processes.
The large value of \MET expected in signal events is due to the weakly interacting LSP particles, which escape without detection.

In the search for an edge, we define two signal regions: either $N_\text{jets} \geq 2$ and $\MET>150\GeV$, or $N_\text{jets} \geq 3$ and $\MET>100\GeV$.
For both regions, we separately consider events in which both leptons satisfy $\abs{\etalep}<1.4$ (``\central'' signal region)
and events in which at least one lepton satisfies $1.6<\abs{\etalep}<2.4$ (``\forward'' signal region).
The motivation for the distinction between the central and forward regions is that for BSM production
through the decay of heavy resonances, the final-state leptons and jets are expected to be more centrally distributed than for the SM backgrounds.
Two methods are used to search for an edge signature.
In the first method, a search for an edge is performed in the range $20 < \mll < 300$\GeV by fitting the signal and background hypotheses to data, as described in Section~\ref{sec:kinfit}.
In the second method, based on a direct comparison of event counts, with no assumption about the shapes of the signal and background distributions,
we select three regions, $20<\mll<70\GeV$, $81<\mll<101\GeV$, and $\mll > 120\GeV$, denoted the ``low-mass'', ``on-Z'', and ``high-mass'' regions, respectively. For this ``counting experiment'',
the integrated yield in each region is compared to the corresponding background prediction.

In the search for BSM  events with an on-shell \Z boson, we perform a dedicated counting experiment in the region $81< \mll <101\GeV$, based on the distribution of \MET.
For this study, two inclusive bins in the number of jets are defined: $N_\text{jets} \geq 2$ and $N_\text{jets} \geq 3$.
Events are examined in exclusive bins of \MET as described in Section~\ref{sec:results}.

While the main SM backgrounds are estimated using data control samples, simulated MC events are used to evaluate smaller sources of background.
The simulation is also used to estimate uncertainties.
Simulated samples of \zjets, \ttbar, $\PW+\text{jets}$, $\mathrm{VV}$, and $\ttbar\mathrm{V}$ ($\mathrm{V}=\PW,\Z$) events
are generated with the \MADGRAPH5.1.3.30 event generator.
The \zjets sample considers events with dilepton invariant masses as low as 10\GeV, as well as decays to the $\tau\tau$ channel.
The matrix element calculation performed with \MADGRAPH5.1.3.30 is then interfaced to the \PYTHIA6.4.22 program for the description of parton showering and
  hadronization, analogous to the procedure used for the signal samples.
The detector response in these samples is simulated with a \GEANTfour model~\cite{Geant} of the CMS detector.
The simulated events are reconstructed and analyzed with the same software used to process data.
In the simulation, multiple pp interactions are superimposed on the hard collision, and the simulated
samples are reweighted to reflect the beam conditions, taking into account the total inelastic pp cross section at the LHC.
Corrections are applied to account for the differences between simulation and data in the trigger and reconstruction efficiencies.
The simulated sample yields are normalized to an integrated luminosity of~\lumifinal\ using
NLO inclusive cross sections, except for the \zjets and \wjets samples, where next-to-next-to-leading order calculations~\cite{Li:2012wna} are used.

\section{Background estimates}\label{sec:bkgd}

The principal SM backgrounds are divided into two categories.
Backgrounds that produce OF pairs $(\Pep\Pgmm, \Pem\Pgmp)$ as often as SF
pairs $(\Pep\Pem, \Pgmp\Pgmm)$ are referred to as flavor-symmetric (FS) backgrounds. This category is dominated by \ttbar processes.
Drell--Yan events form the second principal background category.
The FS background estimate accounts also for $\PW\PW$, $\Zg(\to\tau\tau)$,
and $\mathrm{t}\PW$ single-top quark production, as well as for backgrounds due to leptons from hadron decays and from hadrons misidentified as leptons.
Contributions from $\ttbar + \mathrm{X}$, with $\mathrm{X}$ a \PW, \Z, or Higgs boson,
have been studied in the simulation and found to be negligible.

The missing transverse momentum in \DYjets\ events arises primarily from jet energy resolution and reconstruction effects.
Contributions from SM $\PW\Z$ and $\Z\Z$ processes, which might include genuine \MET, are incorporated into the \DYjets\ background estimates.

\subsection{Flavor-symmetric backgrounds}\label{ssec:flavorSymm}
The contribution of FS background events to the signal regions is
determined using a control sample defined by OF events that satisfy the full event selection.
The OF yields are multiplied by a factor \Rsfof
to account for efficiency differences in the selection  of dilepton pairs in the SF and OF samples.
Two methods are used to evaluate the \Rsfof factor, as explained below.
We also determine corresponding factors \Reeof and \Rmmof for the individual SF terms.

The first method, referred to as the ``factorization'' method, uses a data control sample to determine the SF-to-OF ratio of reconstruction efficiencies,
and a second data control sample to determine the corresponding ratio of trigger efficiencies.
The reconstruction efficiency ratio is evaluated using a large \zjets data control sample selected by requiring 60 $<\mll < $ 120\GeV, $N_\text{jets} \geq 2$, and $\MET<50\GeV$.
The measured quantity is $\rmue = \sqrt{N_{\Pgm\Pgm}/N_{\Pe\Pe}}$,
which represents a ratio of the muon-to-electron reconstruction efficiency uncorrected for trigger efficiencies.
The extrapolation of \rmue into the central (forward) signal region is studied using data.
A 10\% (20\%) systematic uncertainty is assigned to account for the observed dependencies.
The ratio $R_\mathrm{T}$ of the trigger efficiencies in SF and OF events is determined using a data control sample selected with a trigger based on \HT,
which is the scalar sum of jet \pt values for jets with $\pt> 40$\GeV.
The sample is selected by requiring $\HT>200\GeV$, $\mll>20\GeV$, and excluding events with $\njets = 2$ and $\MET > 100\GeV$.
The latter two requirements ensure that this control sample is uncorrelated with the control sample used to determine the reconstruction efficiencies.
A 5\% uncertainty is assigned to the efficiency of each dilepton trigger (\EE, \MM, \EM) to account for the observed dependence of these efficiencies
on \mll, \MET, and $N_\text{jets}$.
The full correction to the OF event rate is 
$\Rsfof = 0.5(\rmue + \rmue^{-1}) R_\mathrm{T}$,
with $R_\mathrm{T} =  {\sqrt{\epsilon^\text{trig}_{\Pe\Pe}\epsilon^\text{trig}_{\mu\mu}} }/{ \epsilon^\text{trig}_{\Pe\mu}}$. 

The second method, referred to as the ``control-region'' method, directly measures the \Rsfof factors from
the ratio of the \EE, \MM, or combined SF yields with the OF yields in a \ttbar-dominated control region with $N_\text{jets} \geq 2$, $100<\MET<150\GeV$,
and $20 < \mll < 70$\GeV.
Differences between the factors obtained in this control region and in the signal regions are studied with the
\ttbar simulation. No difference  is observed within the MC statistical uncertainty, which is used to define the corresponding systematic uncertainty.
The size of this uncertainty is found to be 2\% (3--4\%) in the central (forward) region.

As the two methods rely on uncorrelated control samples, and the results agree within their uncertainties, the final results for \Rsfof are obtained by taking the weighted average of the results from the two methods,
propagating the uncertainties.  The results are shown in Table~\ref{tab:Rsfof}.
The FS background is then given by the number of events in the OF control samples multiplied
by the corresponding \Rsfof factor.
It is seen that the values of the \Rsfof ratios are consistent with unity within the uncertainties.
We find that the variation of \Rsfof for increasing \mll, \njets, and \MET
lies within the assigned uncertainty. We thus use the same value of \Rsfof for all signal regions.

\begin{table}[!htb]
 \renewcommand{\arraystretch}{1.2}
 \begin{center}
  \topcaption{Results for \Rsfof in the signal regions. The results of the two methods are shown with statistical and systematic uncertainties, while the
uncertainties for the combined values are a combination of the statistical and systematic terms. The values of $\rmue$ and $\RT$  listed for the control-region method are not used
directly in the analysis and are listed for purposes of comparison only.}
  \begin{tabular}{l|c|c}
   \hline
   \hline
                           &  Central             & Forward         \\
   \hline

   \multicolumn{3}{c}{Factorization method} \\
   \hline
   $\Rsfof$                          &  $1.03\pm 0.01\pm 0.06$   &  $1.11\pm 0.04\pm 0.08$     \\
   $\Reeof$                    &  $0.47\pm 0.01\pm 0.06$  &  $0.46\pm 0.02\pm 0.10$     \\
   $\Rmmof$                    &  $0.56\pm 0.01\pm 0.07$   &  $0.65\pm 0.03\pm 0.14$     \\

   \hline
   \rmue                             &  $1.09\pm 0.00\pm 0.11$     &  $1.18\pm 0.00\pm 0.24$    \\
   \RT                               &  $1.03\pm 0.01\pm 0.06$    &  $1.10\pm 0.04\pm 0.07$    \\
   \hline

   \multicolumn{3}{c}{Control-region method} \\
   \hline
   $\Rsfof$                          &  $0.99\pm 0.05\pm 0.02$  &  $1.11\pm 0.11\pm 0.03$     \\
   $\Reeof$                        &  $0.44\pm 0.03\pm 0.01$  &  $0.49\pm 0.06\pm 0.02$     \\
   $\Rmmof$                    &  $0.55\pm 0.03\pm 0.01$  &  $0.62\pm 0.07\pm 0.02$     \\
   \hline
   \rmue                             &  $ 1.12\pm 0.04$\stat    & $1.12\pm 0.08 $\stat          \\
   \RT                               &  $0.98\pm 0.05$\stat    & $1.11\pm 0.11$\stat          \\
   \hline

   \hline
   \multicolumn{3}{c}{Combined} \\

   \hline
   \Rsfof                 &  $1.00\pm0.04$        & $1.11\pm0.07$     \\
   \Reeof                 &  $0.45\pm0.03$        & $0.48\pm0.05$     \\
   \Rmmof                 &  $0.55\pm0.03$        & $0.63\pm0.07$     \\
   \hline\hline
 \end{tabular}
 \label{tab:Rsfof}
 \end{center}
\end{table}

\subsection{SM Drell--Yan background}
\label{sec:dybackgrounds}
In the search for an edge based on a fit, the \zjets\ background is determined as described in Section~\ref{sec:kinfit}.
For the counting experiment method, the DY background is determined using the jet-Z balance (JZB)
and \MET-template methods~\cite{zjets2011}, as described in this section.

The \JZB is a measure of the imbalance between the \pt of the \Zg boson
and the \pt of the recoiling hadronic system in \DYjets.
The \JZB is defined as the scalar difference between the \pt of the \Zg and the net \pt of the recoiling hadronic system.
Standard model \DYjets\ events equally populate
negative and positive values of \JZB, because non-zero \JZB in these events arises from jet energy resolution effects, whereas in BSM
and \ttbar events, which contain genuine \MET, \JZB can be very asymmetric towards positive
values because of the correlated production of the lepton pair and the undetected
particles. Events with negative values of \JZB mainly arise from \DYjets processes,
with a small contribution from \ttbar production.
The number of \ttbar events that contribute negative JZB values is determined using the corresponding results for OF events.
The \ttbar contribution is then subtracted from the number of negative JZB events in the SF sample to estimate the
DY+jets background. Uncertainties arising from imperfect knowledge of the \Rsfof factor (Table~\ref{tab:Rsfof}) when subtracting the \ttbar contribution
are propagated to the final \DYjets estimate.
This method also accounts for processes with DY$\to \ell^{\pm}\ell^{\mp} + X$, where $X$
denotes other particles that might be present in the final state.
The systematic uncertainty in the assumption that \DYjets events
equally populate positive and negative values of \JZB is evaluated in simulation by comparing the
\MET distributions of events with $\JZB<0$ and $>$0.
A systematic uncertainty of 20\% is assigned to account for
possible differences, dominated by the limited statistical precision of the MC sample.

The \MET-template method relies on a data control sample consisting of events with photons and jets
to evaluate the \zjets\ background in a high \MET signal region.
For both \DYjets and \gjets events, large values of \MET arise from the mismeasurement of jet \pt values.
The \MET distribution for \DYjets events can thus be evaluated using \gjets events
selected with similar kinematic requirements.
After selection, the \njets, \HT, and boson \pt distributions of the \gjets
sample are reweighted to match those of the \DYjets sample.
The systematic uncertainty is determined in simulation by applying this reweighting to a \gjets MC sample
and comparing the reweighted \MET spectrum to that in a \zjets MC sample.
The uncertainty is taken as the larger of the difference between the two samples or the MC statistical uncertainty.
The assigned uncertainty is 4\% for $\MET < 60\GeV$,
15\% for $60 < \MET < 200\GeV$, 34\% for $200 < \MET < 300\GeV$, and 100\% for $\MET > 300\GeV$.
The uncertainty in the last two \MET bins is mostly due to the MC statistical uncertainty.
Uncertainties are also assigned to account for the difference in the number of pileup interactions in the triggered \zjets and \gjets events
(1--3\%, increasing with \MET) and the purity of the photon selection (1--5\%, increasing with \MET).

With the \MET-template method, the backgrounds from \PW\Z, \ZZ, and other rare SM processes are estimated using
simulation.  An uncertainty of 50\% is assigned to the \PW\Z and \ZZ backgrounds based on comparisons with data in
orthogonal control samples selected by requiring $\njets \geq 2$ and either exactly three leptons (\PW\Z control sample)
or exactly four leptons (\ZZ control sample).  For the other rare backgrounds, which include events with
$\ttbar\Z$, $\Z\Z\Z$, $\Z\Z\PW$, and $\Z\PW\PW$ production, an uncertainty of 50\% is similarly assigned.

The two methods are used to estimate the yield of SM DY events in the on-Z region, with
all other selection criteria the same as for signal events.
Since the two methods use uncorrelated samples to describe the high-\MET tail of SM \DYjets production, the DY background estimate
in the on-Z region is taken to be the weighted average of the two estimates. The individual results are consistent with the weighted average within their uncertainties.

This estimate of the \DYjets background is extrapolated outside on-\Z region using the ratio \Rinout.
It is measured in data as the event yield outside the on-\Z region divided by the yield in the on-\Z region,
for the dilepton invariant mass distribution in SF events with $\MET < 50\GeV$ and $N_{\text{jets}} \geq 2$,
after subtraction of the FS backgrounds.
This distribution is almost entirely composed of lepton pairs from DY processes,
making it suitable for the determination of \Rinout.
A systematic uncertainty of 25\% is assigned to \Rinout to account for possible biases introduced by the different selection criteria used for the signal and control regions,
and for differences in the value of \Rinout between electrons and muons.

\section{Kinematic fit}
\label{sec:kinfit}

The search for an edge based on the fit method is performed using a simultaneous extended unbinned maximum likelihood fit to the dilepton mass distributions of \EE, \MM, and \EM events.
The likelihood model contains three components: (a)~an FS background component, (b)~a DY background component, and (c)~a signal component.

The FS background is described using a model with three regions:
a low-mass region modeling the rising distribution shaped by lepton acceptance requirements,
a transition region, and a high-mass region in which the distribution falls exponentially. The probability density functions describing the three regions are:
\begin{equation}
{\mathcal{P}}_{FS}(m_{\ell\ell}) = \begin{cases} \mathcal{P}_{FS,1}(m_{\ell\ell}) = c_{1} \, m_{\ell\ell}^{\alpha} &\text{if}\quad  20\GeV < m_{\ell\ell} < m_{\ell\ell}^{(1)}, \\
\mathcal{P}_{FS,2}(m_{\ell\ell}) = \sum_{i=1}^{4} c_{2,i} \, m_{\ell\ell}^{i-1} &\text{if}\quad m_{\ell\ell}^{(1)}<m_{\ell\ell}<m_{\ell\ell}^{(2)}, \\
\mathcal{P}_{FS,3}(m_{\ell\ell}) = c_{3}\, \re^{-\beta m_{\ell\ell}} &\text{if}\quad m_{\ell\ell}^{(2)}<m_{\ell\ell}<300\GeV, \\
\end{cases}
\end{equation}
where $m_{\ell\ell}^{(1)}$ and $m_{\ell\ell}^{(2)}$ define the boundaries between the regions. Because of the requirement that the function and its derivative both be continuous,
the FS model is left with five independent parameters plus the normalization.

The DY background is modeled with the sum of an exponential
function, which describes the low-mass rise, and a Breit--Wigner
function with a mean and width set to the nominal \Z boson values~\cite{Beringer:1900zz},
which accounts for the on-\Z lineshape.
To account for the experimental resolution, the Breit--Wigner function is convolved with a double-sided Crystal-Ball~\cite{Crystal} function $\mathcal{P}_{DSCB}(m_{\ell\ell})$:
\begin{equation}
\mathcal{P}_{DSCB}(m_{\ell\ell}) = \begin{cases} A_{1} (B_{1}-\frac{m_{ll}-\mu_{CB}}{\sigma_{CB}})^{-n_{1}} &\text{if}\quad  \frac{m_{ll}-\mu_{CB}}{\sigma_{CB}}<-\alpha_{1}, \\
\exp\left(-\frac{(m_{ll}-\mu_{CB})^2}{2\sigma_{CB}^2}\right) &\text{if}\quad -\alpha_{1}<\frac{m_{ll}-\mu_{CB}}{\sigma_{CB}}<\alpha_{2}, \\
A_{2} (B_{2}+\frac{m_{ll}-\mu_{CB}}{\sigma_{CB}})^{-n_{2}} &\text{if}\quad \frac{m_{ll}-\mu_{CB}}{\sigma_{CB}}>\alpha_{2}, \\
\end{cases}
\end{equation}
where
\begin{equation}
A_{i} = \left(\frac{n_{i}}{|\alpha_{i}|}\right)^{n_{i}} \exp\left(-\frac{\abs{\alpha_{i}}^2}{2}\right) \quad \text{and}\quad B_{i} = \frac{n_{i}}{\abs{\alpha_{i}}}-\abs{\alpha_{i}} .
\end{equation}
The full model for the on-\Z DY lineshape is thus:
\begin{equation}
\mathcal{P}_{DY,\text{~on-Z}} (m_{\ell\ell}) = \int \mathcal{P}_{DSCB}(m_{\ell\ell})\mathcal{P}_{BW}(m_{\ell\ell}-m')\, \rd{}m' .
\end{equation}

The signal component is described by a triangular shape, convolved with a Gaussian distribution to account for the experimental resolution:

\begin{equation}
 {\mathcal{P}}_{S}(m_{\ell\ell}) = \frac{1}{\sqrt{2\pi\sigma_{\ell\ell}}} \int_{0}^{m_{\ell\ell}^\text{edge}} y \, \exp\left( -\frac{(m_{\ell\ell}-y)^2}{2\sigma_{\ell\ell}^{2}}\right) \,\rd{}y.
\end{equation}

As a preliminary step, a fit is performed separately for electrons and muons in the DY-enriched control region (the same control region as described for \Rinout
in Section~\ref{sec:dybackgrounds})
to determine the shape of backgrounds containing a \Z boson.
In systematic studies of the fit, the DY shape parameter that models the exponentially falling spectrum of the virtual photon is varied by $\pm$25\%.
The total effect on the fitted signal yield is found to be negligible.
The parameters of the DY shape are then fixed and only the normalizations of these backgrounds are free parameters in the fit.
The nominal fit is applied simultaneously in the signal region
to the dilepton invariant mass distributions in the \EE, \MM, and \EM~samples.
Therefore the model for the FS background is the same for the SF and OF events.
The \Rsfof factors, in the central and forward regions, are treated as nuisance parameters, parametrized by Gaussian distributions with a mean value and standard deviation given by the values of \Rsfof and their uncertainties (Table~\ref{tab:Rsfof}).
The fit is carried out in the central and forward regions simultaneously,
with the position of the edge as the only shared parameter.
Therefore the signal model has three free parameters: the fitted signal yield in the central and forward regions separately, and the position of the edge.
\section{Results}
\label{sec:results}
\label{ssec:edge}
\label{ssec:cnc}

\begin{figure}[!htbp]
  \begin{center}
  \includegraphics[width=0.47\linewidth]{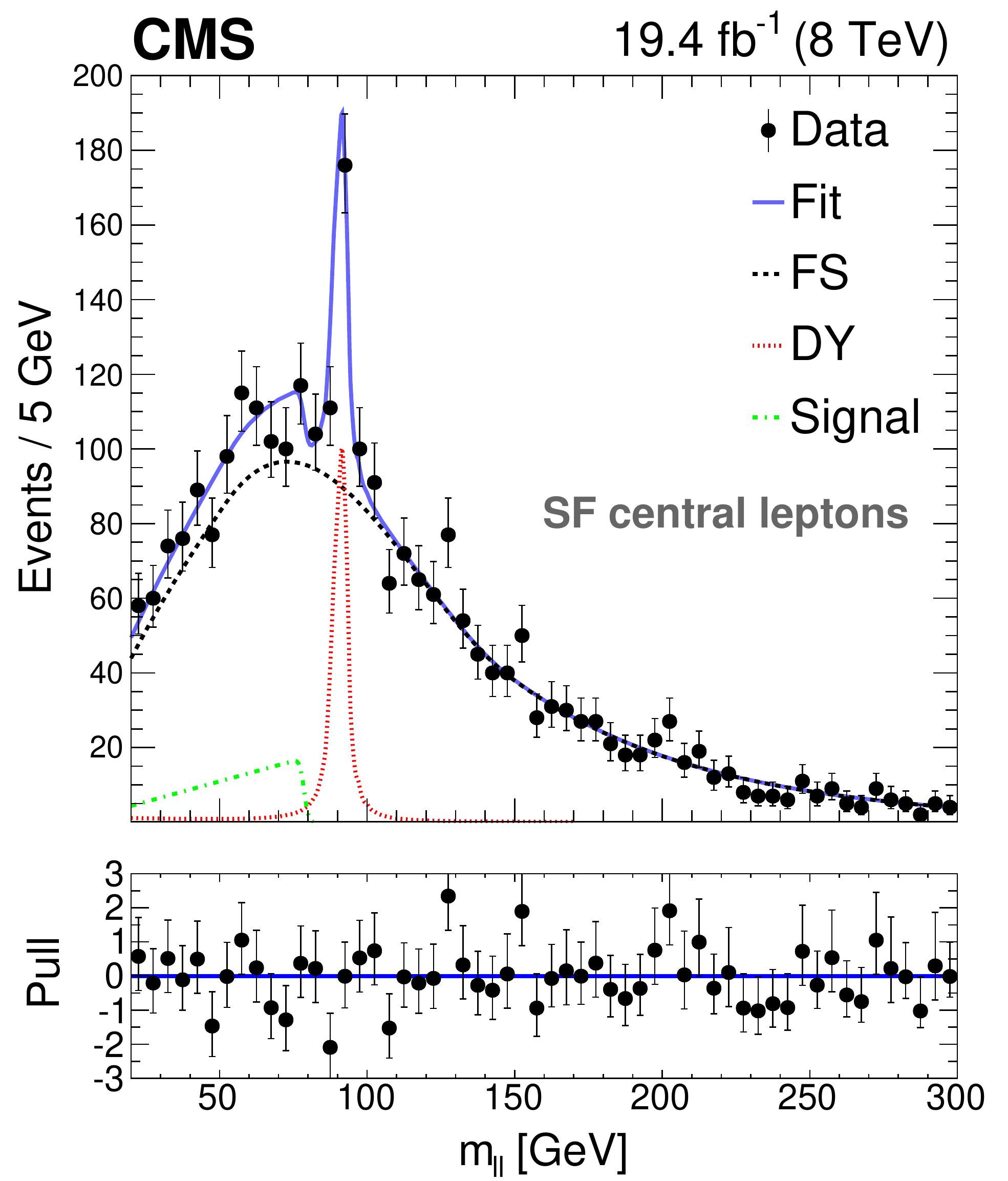}
  \includegraphics[width=0.47\linewidth]{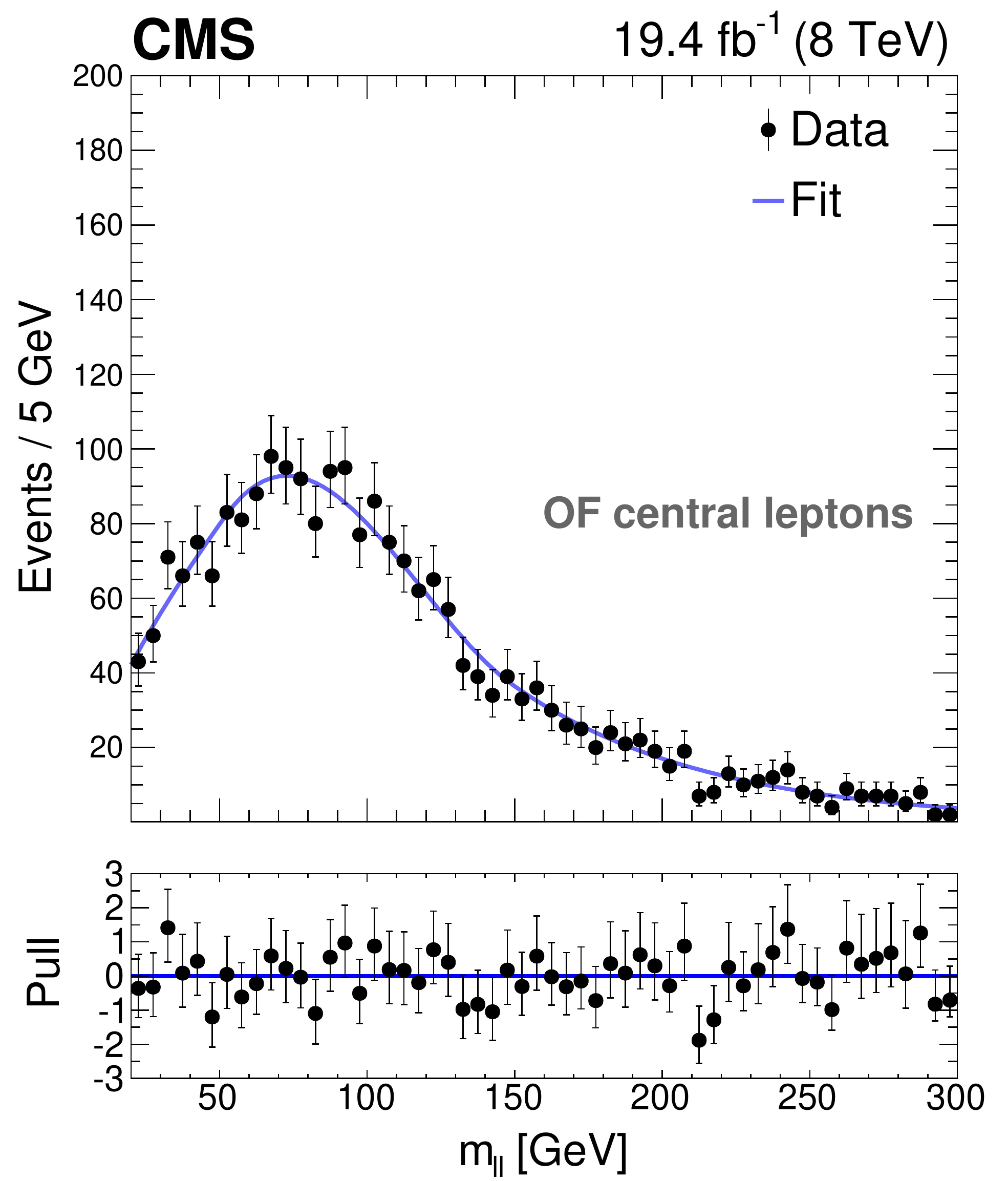}\\
  \includegraphics[width=0.47\linewidth]{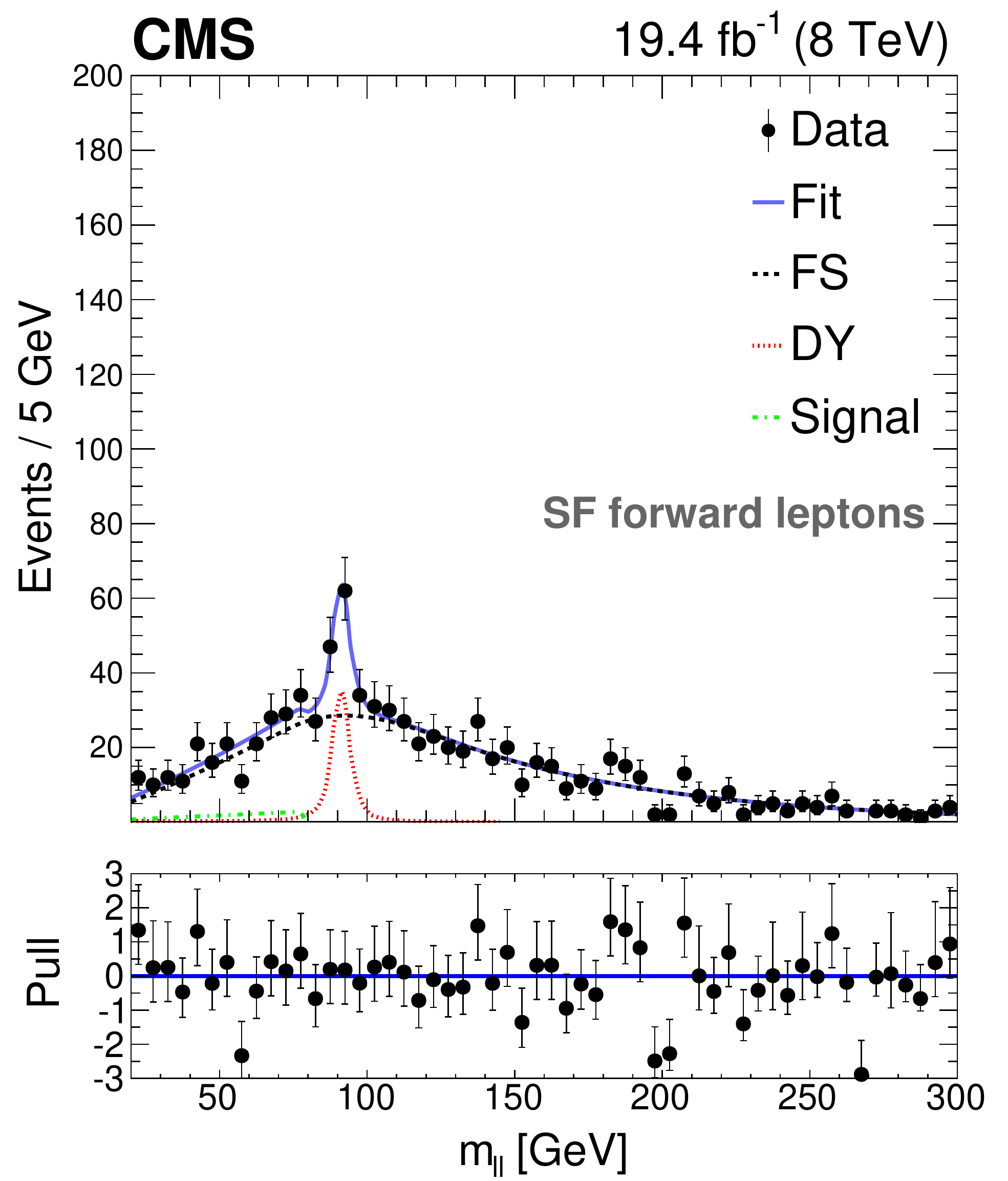}
  \includegraphics[width=0.47\linewidth]{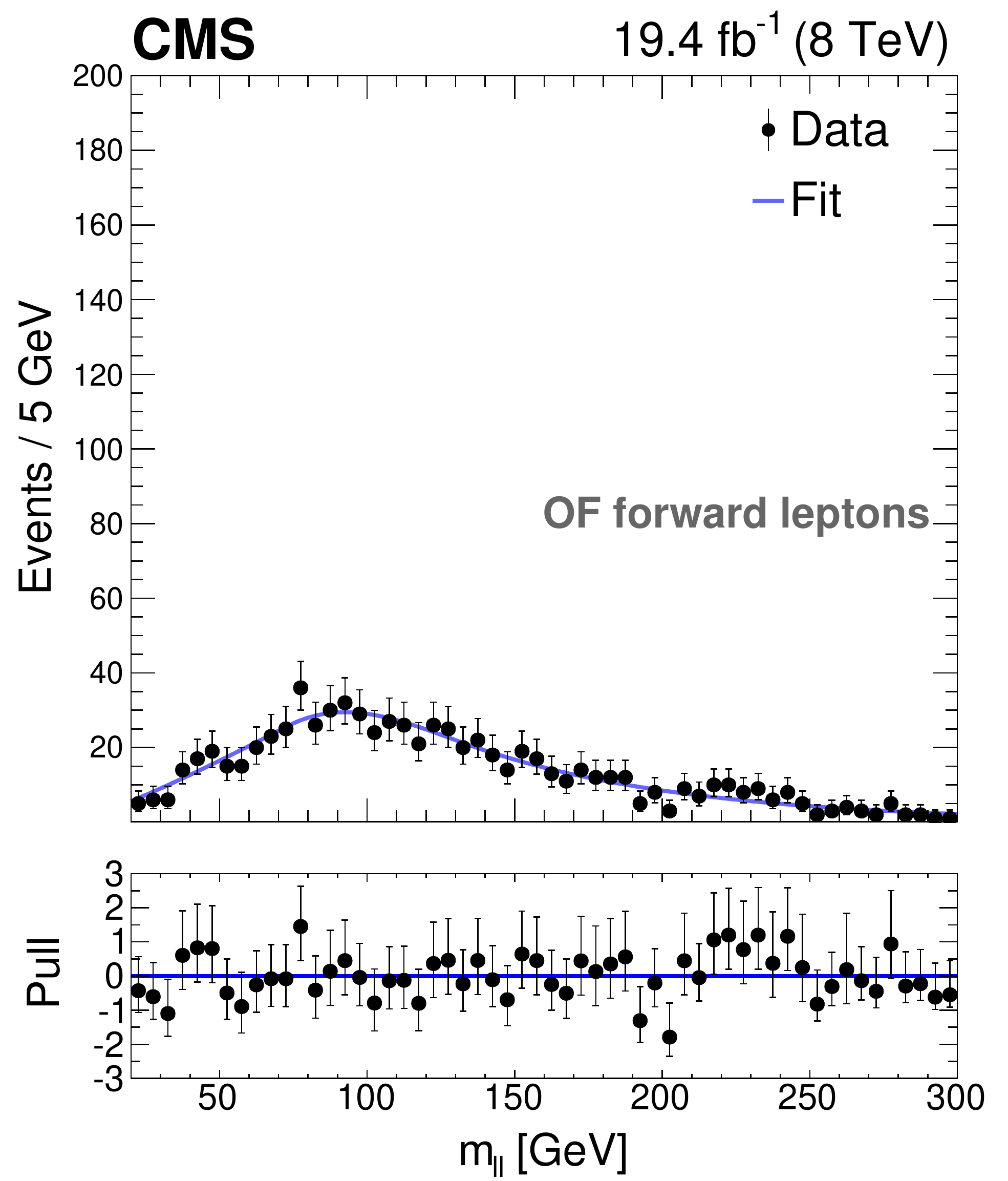}
  \end{center}
  \caption{
Fit results for the signal-plus-background hypothesis in comparison with the measured dilepton mass distributions, in the central (top)
and forward (bottom) regions, projected on the same-flavor (left) and opposite-flavor (right) event samples.
  The combined fit shape is shown as a blue, solid line.
  The individual fit components are indicated by dashed lines.
   The flavor-symmetric (FS) background is displayed with a black dashed line.
   The Drell--Yan (DY) background is displayed with a red dashed line.
   The extracted signal component is displayed with a green dashed line.
    The lower plots show the pull distributions, 
    defined as $(N_\text{data} - N_\text{fit})/\sigma_\text{data}$.
}
  \label{fig:edgefit:fitresultsH1}
\end{figure}

The dilepton mass distributions and the results of the fit in the \central and \forward signal regions are shown in Fig.~\ref{fig:edgefit:fitresultsH1}.
Table~\ref{tab:fitResults} presents a summary of the fit results. A signal yield of $126 \pm 41$ ($22 \pm 20$) events is obtained when evaluating the
signal hypothesis in the \central (\forward) region, with an edge located at $78.7\pm 1.4\GeV$.
The p-value, evaluated using $-2\ln Q$, is 0.009, where $Q$ denotes the ratio of the fitted likelihood value for the
signal-plus-background hypothesis to the background-only hypothesis for the case where the edge position is fixed to the observed value. This p-value is interpreted as the one-sided tail probability of a Gaussian distribution and corresponds to an excess in the observed number of events compared to the SM background estimate of 2.4 standard deviations.

As cross-checks, we tested alternative shapes for the FS background, specifically the sum of three Gaussian distributions, and binned and smoothed histograms taken from the OF events. In all cases, the results were found to be consistent with the nominal results.

\begin{table}[hbtp]
 \renewcommand{\arraystretch}{1.3}
 \centering
 \topcaption{Results of the unbinned maximum likelihood fit for event yields in the signal regions.
The quoted uncertainties are calculated using the MINOS~\cite{James:1975dr} program and account for both statistical and systematic sources.}
  \label{tab:fitResults}
  \begin{tabular}{l| cc}
    \hline
    \hline
                                             &  Central               &  Forward        \\
    \hline
             Drell--Yan                      & $158 \pm 23$            & $71 \pm 15$       \\
             OF yield                & $2270 \pm 44$           & $745 \pm 25$          \\
             \Rsfof                          & $1.03 \pm 0.03$                 & $1.02 \pm 0.05$             \\
             Signal events                   & $126 \pm 41$            & $22 \pm 20$         \\
             $\mll^\text{edge} $         & \multicolumn{2}{c}{$78.7\pm 1.4$\GeV}  \\
    \hline
  Local significance                   & \multicolumn{2}{c}{2.4\,$\sigma$}           \\
    \hline
    \hline
  \end{tabular}

\end{table}

\begin{figure}[hbtp]
  \begin{center}
   \includegraphics[width=0.49\textwidth]{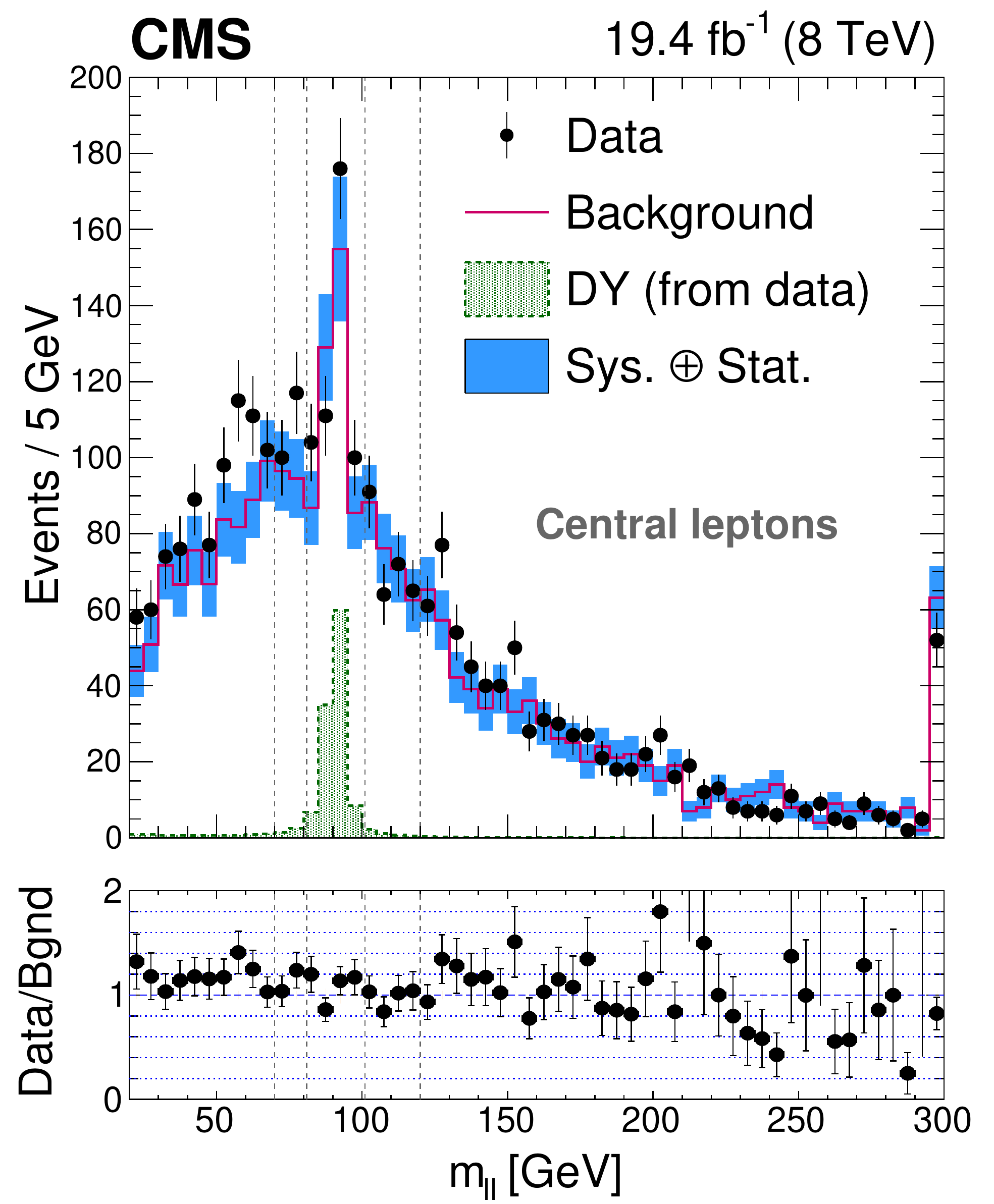}
   \includegraphics[width=0.49\textwidth]{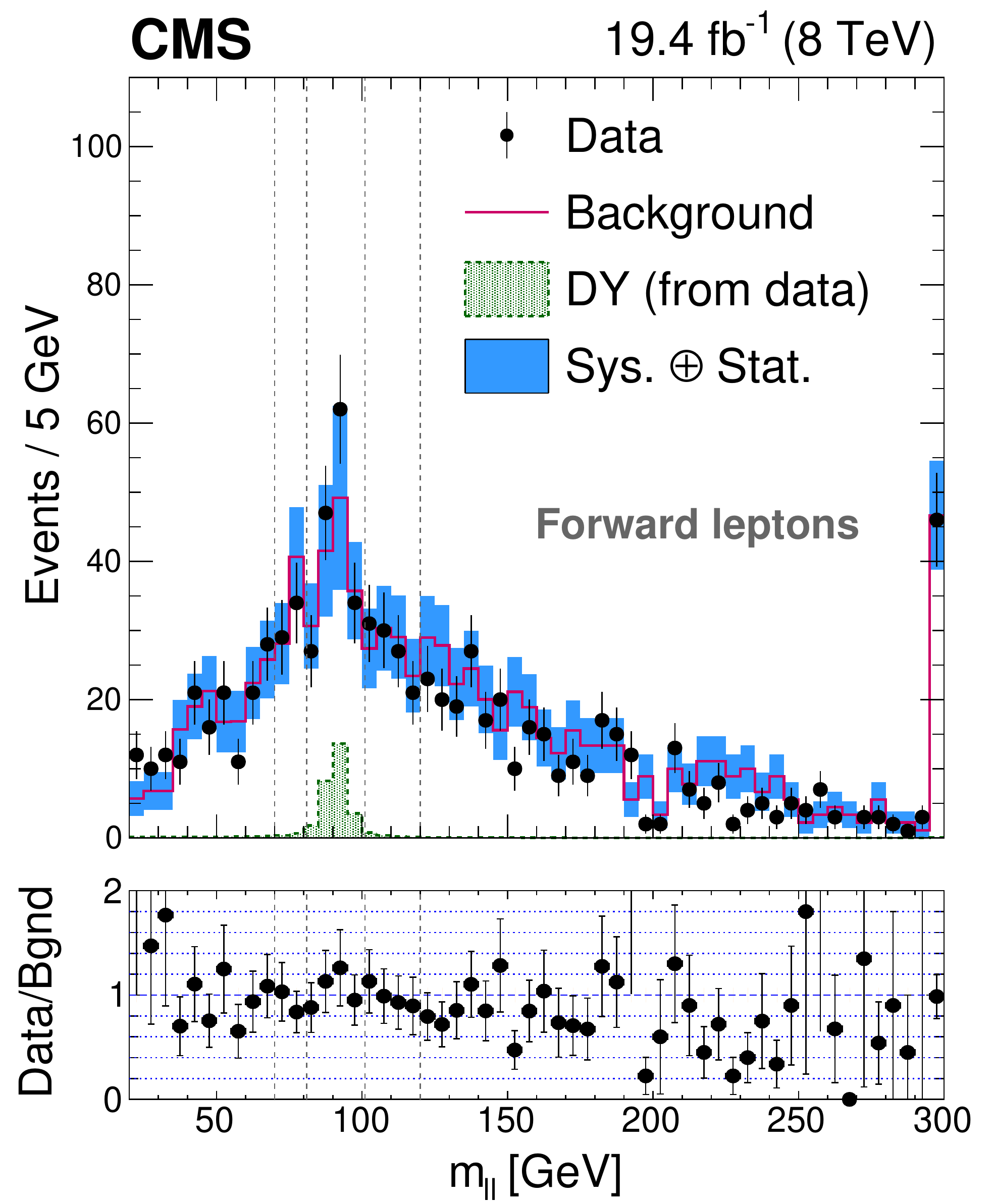}
    \caption{Comparison between the observed and estimated SM background dilepton mass distributions in the (left) \central
    and (right) \forward regions, where the SM backgrounds are evaluated from control samples (see text) rather than from a fit. The rightmost bins contain the overflow.
    The vertical dashed lines denote the boundaries of the low-mass, on-Z, and high-mass regions.
    The lower plots show the ratio of the data to the predicted background.
    The error bars for both the main and lower plots include both statistical and
    systematic uncertainties.
}
    \label{fig:results}
  \end{center}
\end{figure}

Besides the fit described in Section~\ref{sec:kinfit}, we perform a counting experiment in the mass windows $20<\mll<70\GeV$, $81<\mll<101\GeV$, and $\mll>120\GeV$, with no assumption about a particular signal shape, as mentioned in Section~\ref{sec:eventsel}.
Figure~\ref{fig:results} shows the invariant mass distributions for the signal candidate sample and the estimated background. For the background prediction, the OF yield in the signal mass window is multiplied by the \Rsfof factor, and the background prediction for backgrounds containing a \Z boson in the on-Z region by the \Rinout factor,
as described in Sections~\ref{ssec:flavorSymm} and~\ref{sec:dybackgrounds}.

The results are summarized in Table~\ref{tab:METresults2012}. The significance of the excess in the observed number of events with respect to the estimated number of SM background
events is evaluated using a profile likelihood asymptotic approximation~\cite{HiggsTool1}.
The local significance of the excess in the central low-mass region, where the largest deviation is found, is 2.6 standard deviations.
Note that the signal regions were defined before the data sample was examined, and that the low-mass region ($20<\mll<70$\GeV) does not include events between
70\GeV and the best-fit value for the location of the edge ($\mll=78.7$\GeV).
The flavor of the $\ell^+\ell^-$ pair was studied in the counting experiment.
Within the statistical uncertainty and accounting for differences in the reconstruction efficiencies, the electron and the muon channels are found to contribute evenly to the excess.
Further studies of the excess in the low-mass region do not yield evidence for a neglected systematic term. The excess is observed predominantly
in events with at least one identified bottom quark jet (b jet) and diminishes if a veto on the presence of a b jet is applied.
To identify b jets, we use the CSV algorithm at the medium working point~\cite{Chatrchyan:2012jua}.

\begin{figure}[hbtp]
\begin{center}
  \includegraphics[width=0.49\textwidth]{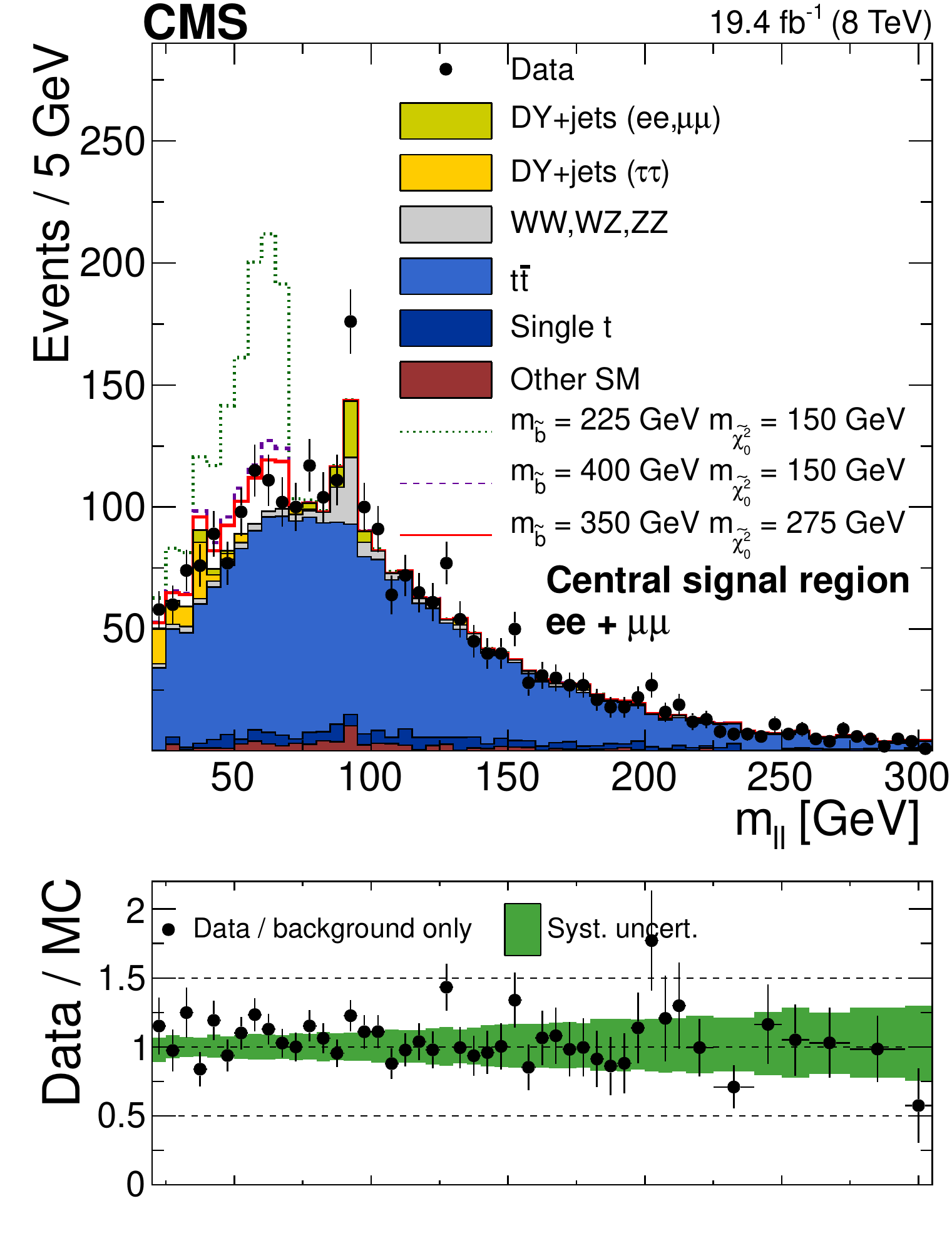}
  \includegraphics[width=0.49\textwidth]{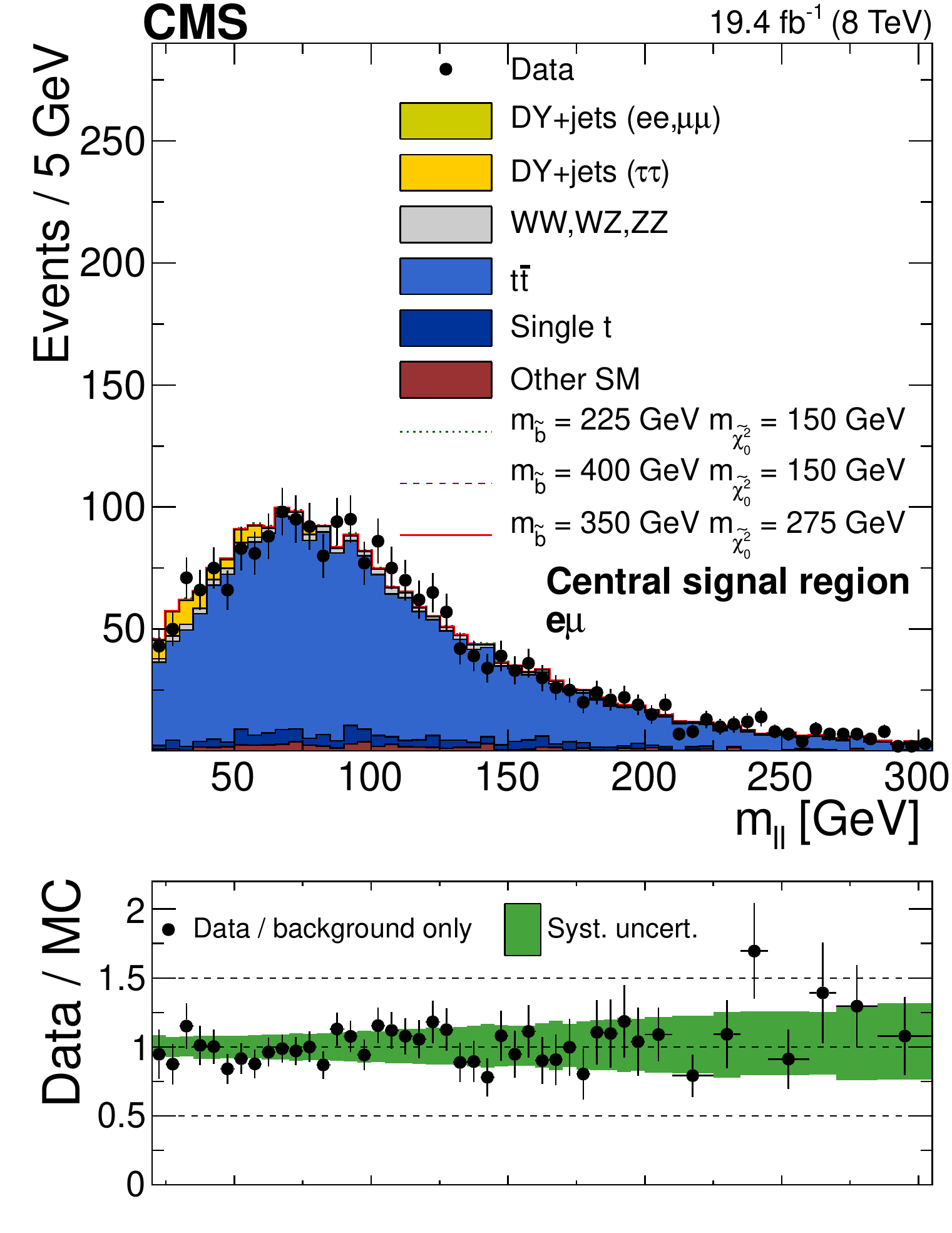}\\
  \caption{Data compared with SM simulation for the SF (left) and OF (right) event samples in the central region.
Example signal scenarios based on the pair production of bottom squarks are shown (see text).
In the ratio panel below each plot, the error bars on the points show the statistical uncertainties of the data and MC samples, while the shaded band
indicates the MC systematic uncertainty.
}
  \label{fig:dataMC_signal}
\end{center}
\end{figure}

Figure~\ref{fig:dataMC_signal} presents a comparison of the data and the SM simulation in the central region.
Expectations for the fixed-edge bottom-squark pair-production scenario of Fig.~\ref{fig:feynmanEdge}~(left), with mass combinations ($m_{\PSQb}$,$m_{\secondchi}$) = (225, 150)\GeV, (350, 275)\GeV, and (400, 150)\GeV for the bottom squark and \secondchi, are also shown.
The first scenario presents the illustration of a model that can easily be excluded,
while the other two present examples of models that are consistent with our data.

\begin{table}[hbtp]
 \renewcommand{\arraystretch}{1.3}
 \centering
 \topcaption{Results of the edge-search counting experiment for event yields in the signal regions.
     The statistical and systematic uncertainties are added in quadrature, except for the flavor-symmetric backgrounds.
     Low-mass refers to $20 < \mll < 70$\GeV, on-\Z to  $81 < \mll < 101$\GeV, and high-mass to $\mll > 120$\GeV.
     }
  \label{tab:METresults2012}
  \resizebox{\textwidth}{!}{
  \begin{tabular}{l| cc | cc | cc}
    \hline
    \hline
    							& \multicolumn{2}{c}{Low-mass} & \multicolumn{2}{c}{On-\Z} & \multicolumn{2}{c}{High-mass} \\
    \hline
                                &  Central        & Forward  &  Central  & Forward   &  Central        & Forward \\
    \hline
            Observed       &  860                   & 163              &  487            &  170       &   818           &   368    \\
    \hline
             Flavor-symmetric    & $722\pm27\pm29$        & $155\pm13\pm10$  &  $355\pm19\pm14$ & $131\pm12\pm8$ & $768\pm28\pm31$ & $430\pm22\pm27$ \\
             Drell--Yan          & $8.2\pm2.6$            & $2.5\pm1.0$      & $116\pm21$ & $42\pm9$ & $2.5\pm0.8$ & $1.1\pm0.4$ \\
    \hline
             Total estimated    & $730\pm40$             & $158\pm16$       & $471\pm32$ & $173\pm17$ & $771\pm42$ & $431\pm35$ \\
    \hline
         Observed$-$estimated  & $130^{+48}_{-49}$      & $5^{+20}_{-20}$ & $16^{+37}_{-38} $ & $-3^{+20}_{-21}$ & $47^{+49}_{-50}$ & $-62^{+37}_{-39} $ \\
    \hline
   Significance      & \significanceNumber\,$\sigma$    &  0.3\,$\sigma$  & 0.4\,$\sigma$ & $<$0.1\,$\sigma$ &0.9\,$\sigma$ & $<$0.1\,$\sigma$ \\
   \hline
    \hline
  \end{tabular}
}
\end{table}

\begin{figure}[!ht]
  \begin{center}
      \includegraphics[width=0.49\linewidth]{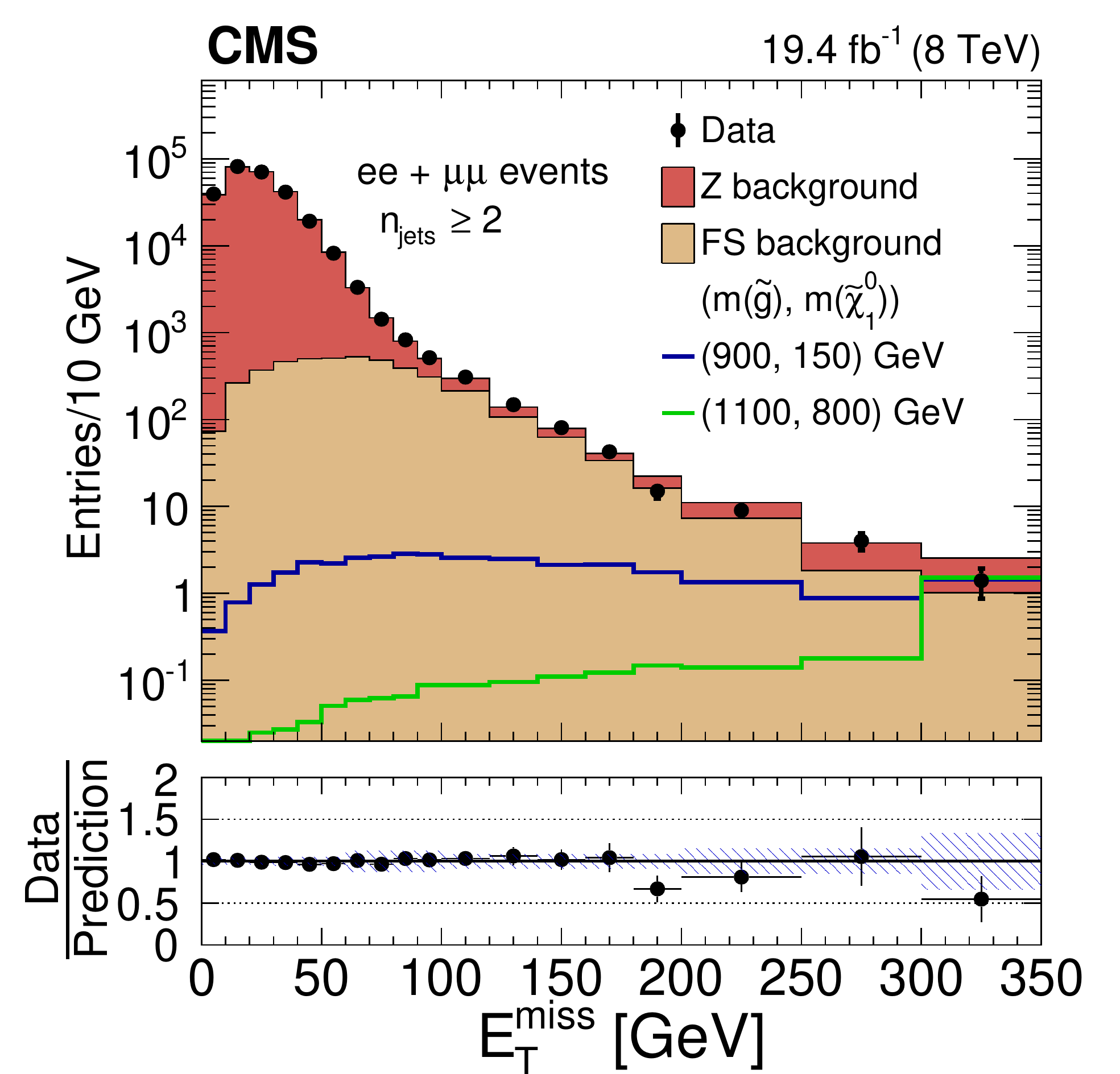}
      \includegraphics[width=0.49\linewidth]{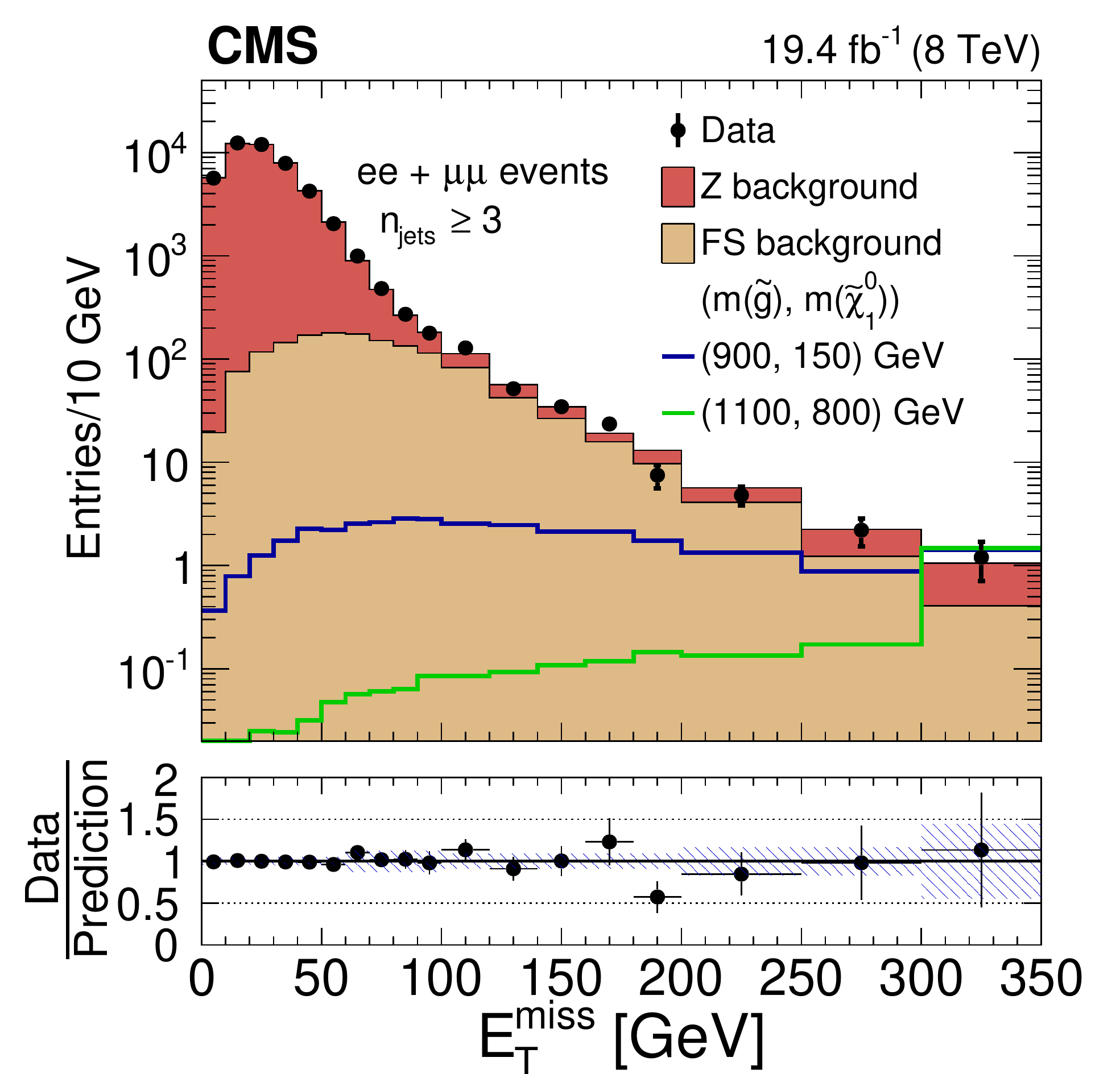}
    \caption{\label{fig:results_onZ}
      The \MET\ distributions for the on-\Z signal regions with (left) $\geq$2~jets and (right) $\geq$3~jets.
      The uncertainty band shown for the ratio includes both statistical and systematic uncertainties.
      Additionally, \MET\ distributions are drawn for two choices of masses in the GMSB scenario.
      The rightmost bins contain the overflow.
    }
  \end{center}
\end{figure}

The results from the dedicated on-Z counting experiment mentioned in Section~\ref{sec:eventsel} are presented in Tables~\ref{tab:results_onZ_2jets}
and~\ref{tab:results_onZ_3jets} for events with $N_\text{jets} \geq 2$ and $N_\text{jets} \geq 3$, respectively.
The corresponding \MET distributions are shown in Fig.~\ref{fig:results_onZ}.
The data are seen to agree with the SM predictions
across the full \MET spectrum.

\begin{table}[htb]
\begin{center}
\topcaption{\label{tab:results_onZ_2jets}
Event yields in the signal region for the dedicated on-\Z counting experiment with $N_\text{jets} \geq 2$. Both statistical and systematic uncertainties are included
for the background estimates.  Signal yields are also shown for two choices of masses (in \GeVns{}) in the GMSB scenario (statistical uncertainties only).
}
\begin{tabular}{l|c|c|c}
\hline
\hline
         \MET~(\GeVns{})       & 100--200 & 200--300 & $>$300  \\
\hline
          DY background &$ 336 \pm 89 $&$ 28.6 \pm 8.6 $&$ 7.7 \pm 3.6   $\\
         FS background &$ 868 \pm 57 $&$ 45.9 \pm 7.3 $&$ 5.1 \pm 2.3   $\\
\hline
      Total background &$ 1204 \pm 106 $&$ 74.5 \pm 11.3 $&$ 12.8 \pm 4.3   $\\
           Data &           1187 &           65 &              7  \\
\hline
\multicolumn{4}{c}{GMSB signal yields}\\
\hline
$m_{\PSg} = 900$, $m_{\PSGczDo} = 150$    &$  22.1 \pm 0.4 $&$  11.1 \pm 0.3  $&$ 7.2 \pm 0.2  $\\
$m_{\PSg} = 1100$, $m_{\PSGczDo} = 800$    &$  1.1 \pm 0.04 $&$  1.6 \pm 0.05  $&$ 7.6 \pm 0.1  $\\
\hline
\hline
\end{tabular}
\end{center}
\end{table}

\begin{table}[htb]
\begin{center}
\topcaption{\label{tab:results_onZ_3jets}
Event yields in the signal region for the dedicated on-\Z counting experiment with $N_\text{jets} \geq 3$. Both statistical and systematic uncertainties are included
for the background estimates. Signal yields are also shown for two choices of masses (in \GeVns{}) in the GMSB (statistical uncertainties only).
}
\begin{tabular}{l|c|c|c}
\hline
\hline
          \MET~(\GeVns{})      & 100--200 & 200--300 & $>$300  \\
\hline
          DY background &$ 124 \pm 33 $&$ 12.7 \pm 3.8 $&$ 3.2 \pm 1.8   $\\
         FS background &$ 354 \pm 28 $&$ 26.5 \pm 5.4 $&$ 2.0 \pm 1.4   $\\
\hline
      Total background &$ 478 \pm 43 $&$ 39.2 \pm 6.6 $&$ 5.3 \pm 2.3   $\\
           Data &           490 &           35 &              6  \\
\hline
\multicolumn{4}{c}{GMSB signal yields}\\
\hline
$m_{\PSg}$ = 900, $m_{\PSGczDo} = 150$    &$  22.0 \pm 0.4 $&$  11.0 \pm 0.3  $&$ 7.1 \pm 0.2  $\\
$m_{\PSg}$ = 1100, $m_{\PSGczDo} = 800$    &$  1.1 \pm 0.04 $&$  1.5 \pm 0.05  $&$ 7.4 \pm 0.1  $\\
\hline
\hline
\end{tabular}
\end{center}
\end{table}

\section{Uncertainties in signal modeling}
\label{sec:systematics}
Systematic uncertainties associated with the estimation of the SM background are discussed in Section~\ref{sec:bkgd}. This section describes uncertainties in the signal modeling. The impact of the uncertainties on the considered signal models is shown in Table~\ref{tab:sysUncerts}.

The uncertainty related to the measurement of the integrated luminosity is 2.6\%~\cite{CMS-PAS-LUM-13-001}.
The uncertainty related to the parton distribution functions (PDF) is evaluated using the PDF4LHC recommendations~\cite{Alekhin:2011sk,Botje:2011sn,Ball:2012cx,Martin:2009iq,Lai:2010vv} and amounts to 0--6\% in the signal acceptance.
Uncertainties in the modeling of the reconstruction, identification, and isolation efficiencies amount to 1\% per lepton. The uncertainty of the corrections to account for lepton reconstruction differences between the fast and full simulations is 1\% per lepton. The uncertainty in the trigger efficiencies is about 5\%. Uncertainties in the muon momentum scale are negligible, whereas for electrons the uncertainty in the energy scale is 0.6\%~(1.5\%) in the barrel (endcap) region, leading to a 0--5\% uncertainty in the signal yield.  The uncertainty in the jet energy scale~\cite{1748-0221-6-11-P11002} varies between 0--8\%. The uncertainty in the jet energy scale is propagated to evaluate the uncertainty associated with the \MET distribution, which is found to be 0--8\%.
The uncertainty associated with the modeling of initial-state radiation (ISR)~\cite{Chatrchyan:2013xna} is 0--14\%.
The uncertainty in the correction to account for pileup in the simulation is evaluated by shifting the inelastic cross section by $\pm5\%$.
The impact on the signal yield is found to be about 1\%.

\begin{table}
\begin{center}
\topcaption{Summary of systematic uncertainties for the signal efficiency.}
\label{tab:sysUncerts}
\begin{tabular}{l|c}
\hline \hline
Uncertainty source & Impact on signal yield [\%]\\ \hline
Luminosity & 2.6 \\
PDFs on acceptance & 0--6 \\
Lepton identification/isolation & 2\\
Fast simulation lepton identification/isolation & 2 \\
Dilepton trigger & 5 \\
Lepton energy scale & 0--5  \\
\MET & 0--8  \\
Jet energy scale/resolution & 0--8  \\
ISR modeling & 0--14 \\
Additional interactions & 1 \\
\hline
\hline
\end{tabular}
\end{center}
\end{table}
\section{Interpretation}
\label{sec:interpretation}

Based on the results of the counting experiments, exclusion limits are determined.
The kinematic fit is not used for this purpose because it assumes a specific shape for the dilepton mass spectrum of signal events and so is more model dependent.
The limits are calculated at 95\% confidence level (CL) using the $\mathrm{CL}_\mathrm{s}$ criterion with the LHC-style test statistic~\cite{HiggsTool1,Junk:1999kv,0954-3899-28-10-313}, taking into account the statistical and systematic uncertainties in the signal yields and the background predictions discussed in Sections \ref{sec:systematics} and \ref{sec:bkgd}, respectively. The different systematic uncertainties are considered to be uncorrelated with each other, but fully correlated among the different signal regions. The systematic uncertainties are treated as nuisance parameters and are parametrized with log-normal distributions.

\begin{figure}[hbtp]
\begin{center}
  \includegraphics[width=0.49\textwidth]{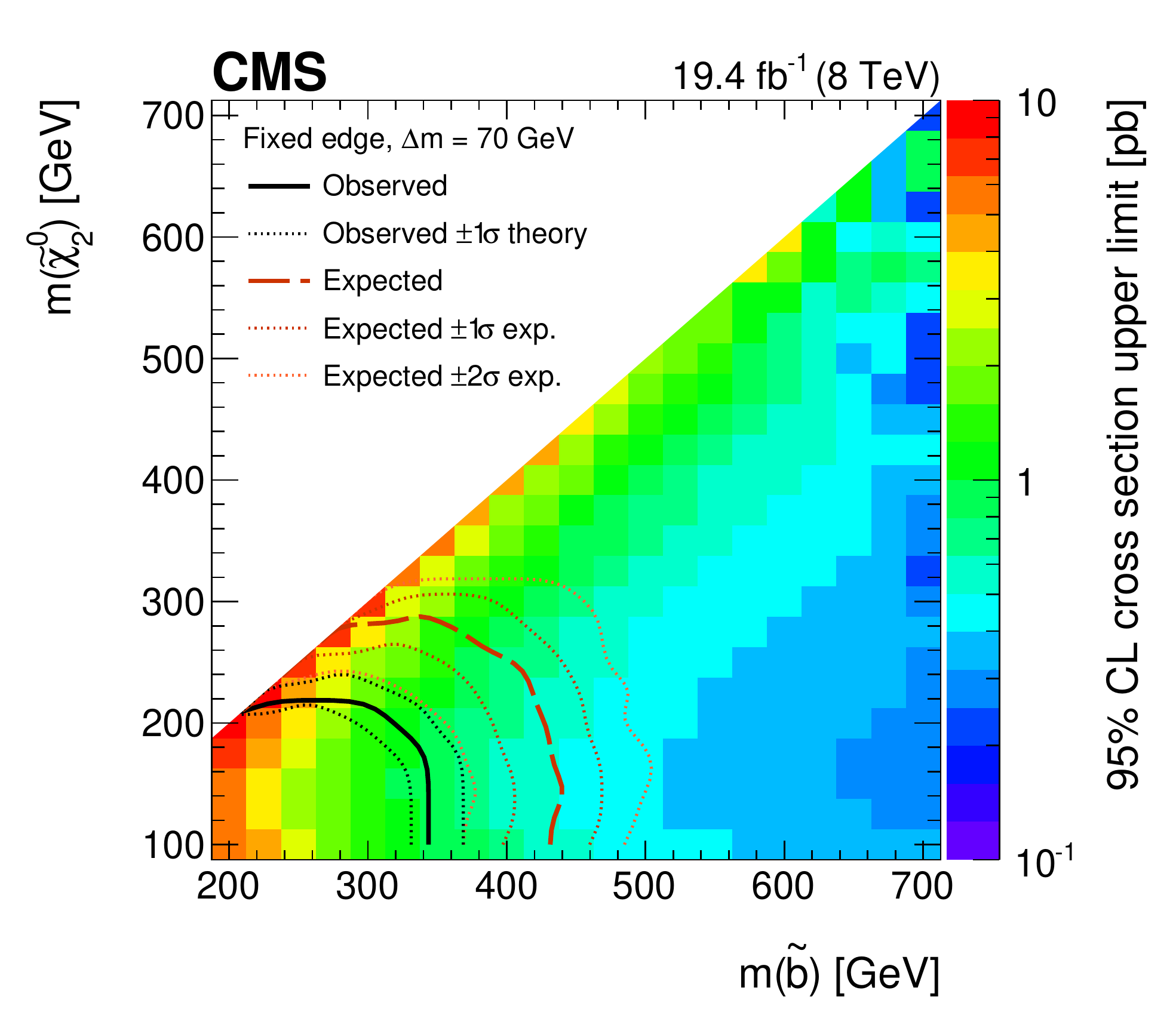}
  \includegraphics[width=0.49\textwidth]{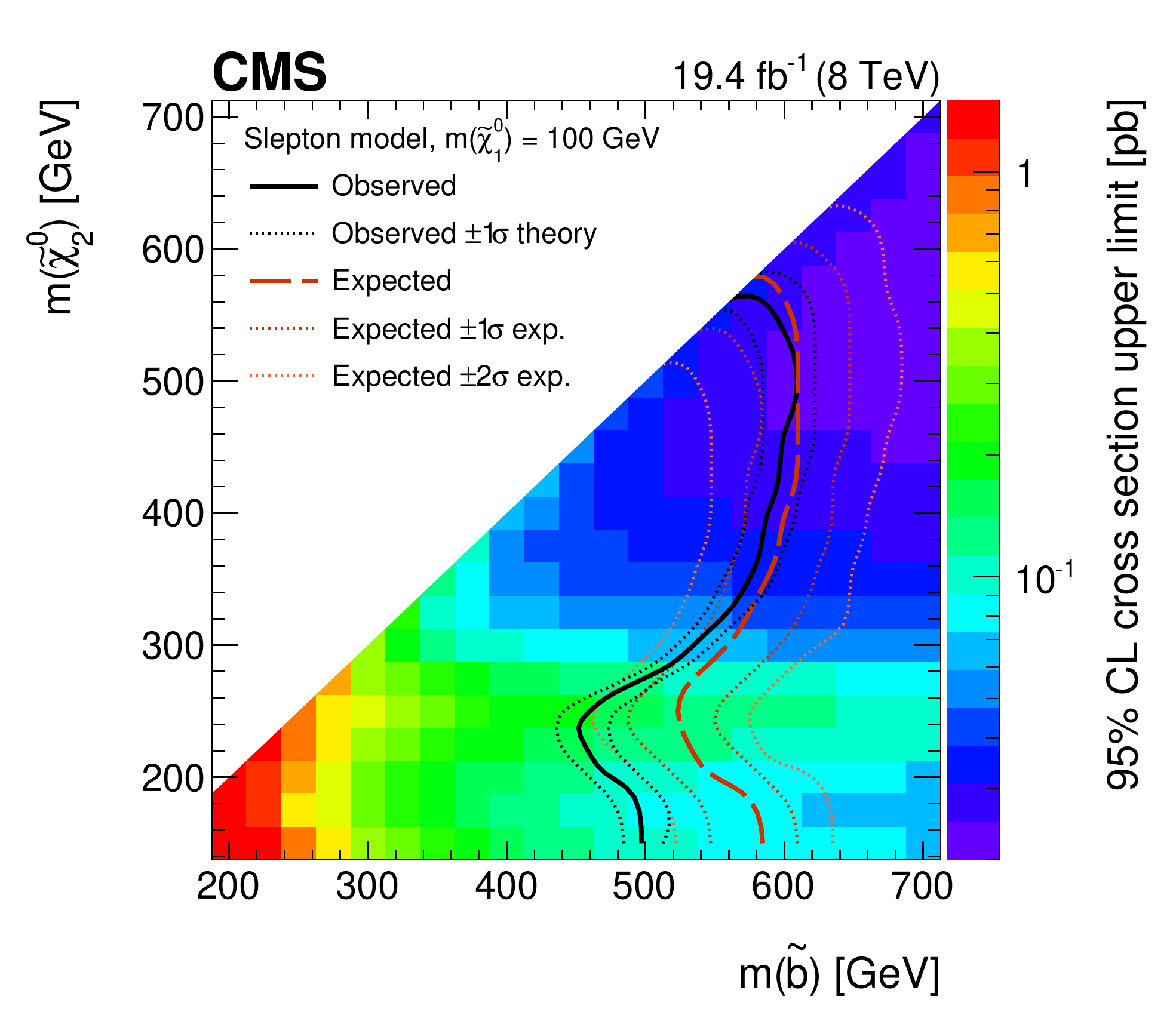}\\
  \caption{Exclusion limits at 95\% CL for the fixed- (left) and slepton-edge (right) scenarios in the $m_{\sbottom}$-$m_{\secondchi}$ plane. The color indicates the excluded cross section for each considered point in parameter space. The intersections of the theoretical cross section with the expected and observed limits are indicated by the solid and hatched lines. The 1 standard deviation ($\sigma$) experimental and theoretical uncertainty contours are shown as dotted lines.}
  \label{fig:limitsT6bbEdge}
\end{center}
\end{figure}

For the fixed- and slepton-edge scenarios all six signal regions (low-mass, on-Z, and high-mass, for the central and forward lepton regions) are combined. 
The resulting exclusion limits for these scenarios are shown in Fig.~\ref{fig:limitsT6bbEdge}. 
The production of bottom squarks is considered, as the excess observed in data consists predominantly of events with at least one b jet. 
In the fixed-edge scenario, the mass difference between the \secondchi and \firstchi neutralinos is fixed to 70\GeV, resulting in an edge in the \mll spectrum at this value. 
This choice for the edge position is motivated by the observed excess in the low-mass region of the counting experiment. 
Bottom-squark masses between 200 and 350\GeV are probed, depending on the value of the \secondchi mass. 
In the slepton-edge scenario, the \firstchi mass is set to 100\GeV, leaving the position of the edge as a free parameter that approximately corresponds to the mass 
difference between the \secondchi and $\PSGczDo$. 
The branching fraction into dilepton final states is larger than in the fixed-edge scenario, and bottom-squark masses between 450 and 600\GeV are probed, again depending 
on the value of the \secondchi mass. The loss of sensitivity seen in Fig.~\ref{fig:limitsT6bbEdge} (right) for \secondchi masses 
around 250\GeV occurs because the peak of the triangular signal shape lies within, or close to, the gaps in acceptance 
between the low-mass and on-Z, or the on-Z and high-mass signal regions.

\begin{figure}[hbtp]
\begin{center}
  \includegraphics[width=0.49\textwidth]{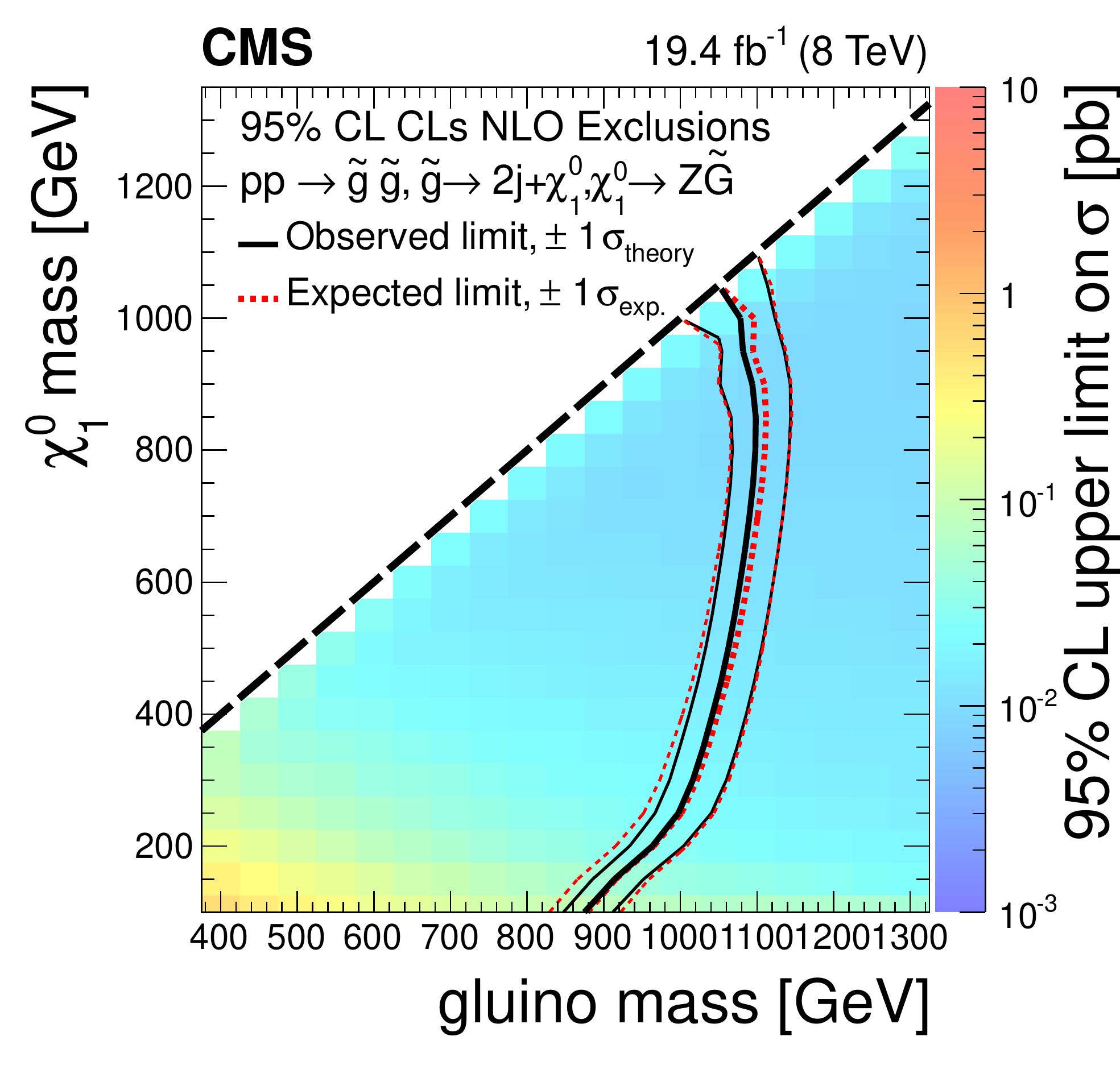}
  \caption{Exclusion limits at 95\% CL for the GMSB scenario in the $m_{\PSg}$-$m_{\firstchi}$ plane. The color indicates the excluded cross section for each considered point in parameter space. The intersections of the theoretical cross section with the expected and observed limits are indicated by the solid and hatched lines. The 1 standard deviation ($\sigma$) experimental and theoretical uncertainty contours are shown as dotted lines.}
  \label{fig:limitsT5ZZgmsb}
\end{center}
\end{figure}

The results from the dedicated on-Z signal regions are used to place limits on the GMSB scenario.
In this scenario, there are two free parameters:
the masses of the gluino ($m_{\PSg}$) and the $\firstchi$ ($m_{\firstchi}$).
As signal events typically have large jet multiplicities, the exclusive bins requiring
$N_\text{jets} \geq 3$ and \MET in the ranges $100 < \MET < 200\GeV$, $200 < \MET < 300\GeV$, and $\MET > 300\GeV$ are used.
The results are shown in Fig.~\ref{fig:limitsT5ZZgmsb} in the plane of $m_{\firstchi}$ versus $m_{\PSg}$.  These results probe gluino masses up to about 900--1100\GeV depending on the \firstchi mass.
The limit is least stringent when $m_{\firstchi}$ is close to the \Z boson mass.

\section{Summary}
\label{sec:conclusion}
  We have presented a search for physics beyond the standard model in
  the opposite-sign dilepton final state using a data sample of
  pp collisions collected at a center-of-mass energy of 8\TeV,
  corresponding to an integrated luminosity of \lumifinal, recorded with
  the CMS detector in 2012.  Searches are performed for signals that either produce a kinematic edge, or a peak at the \Z boson mass, in the dilepton invariant mass distribution. For regions dominated by flavor-symmetric backgrounds,
  i.e., backgrounds that produce opposite-flavor events $(\Pep\Pgmm, \Pem\Pgmp)$ as often as
same-flavor events $(\Pep\Pem, \Pgmp\Pgmm)$, we achieve a precision of about 5\% (10\%) for the estimated
  number of standard model background events in the central (forward) lepton rapidity regions.
We do not observe evidence for a statistically significant signal.
The maximum deviation from the null hypothesis is at the level of 2.6 standard deviations and is observed in the dilepton mass window $20 < \mll< 70$\GeV.

We interpret the results of the search for a kinematic edge in the context of simplified models consisting of bottom-squark pair production, with
each bottom-squark \sbottom decaying to a bottom quark and the \secondchi neutralino.
Exclusion limits are set in the $m_{\sbottom}$-$m_{\secondchi}$ mass plane for two scenarios.
In the fixed-edge scenario,
the mass difference between the \secondchi and \firstchi particles is fixed to 70\GeV. Bottom-squark masses between
200 and 350\GeV are excluded at 95\% confidence level in this scenario.  In the slepton-edge scenario, the \secondchi decays to an on- or off-shell \Z boson and
the \firstchi lightest supersymmetric particle or to a slepton
and a lepton, with a 50\% probability for each possibility.  Bottom-squark masses between 450 and 600\GeV
are excluded at 95\% confidence level in this second scenario. In both scenarios, the sensitivity depends on the mass of the \secondchi.
Finally, a dedicated search for events containing an on-shell \Z boson is interpreted in a model
of gauge-mediated supersymmetry breaking, in which the \Z bosons are produced in decay chains initiated through gluino pair production.
Gluino masses between 900 and 1100\GeV are excluded at 95\% confidence level, depending on the mass of the lightest neutralino \firstchi.

\begin{acknowledgments}
We congratulate our colleagues in the CERN accelerator departments for the excellent performance of the LHC and thank the technical and administrative staffs at CERN and at other CMS institutes for their contributions to the success of the CMS effort. In addition, we gratefully acknowledge the computing centers and personnel of the Worldwide LHC Computing Grid for delivering so effectively the computing infrastructure essential to our analyses. Finally, we acknowledge the enduring support for the construction and operation of the LHC and the CMS detector provided by the following funding agencies: BMWFW and FWF (Austria); FNRS and FWO (Belgium); CNPq, CAPES, FAPERJ, and FAPESP (Brazil); MES (Bulgaria); CERN; CAS, MoST, and NSFC (China); COLCIENCIAS (Colombia); MSES and CSF (Croatia); RPF (Cyprus); MoER, ERC IUT and ERDF (Estonia); Academy of Finland, MEC, and HIP (Finland); CEA and CNRS/IN2P3 (France); BMBF, DFG, and HGF (Germany); GSRT (Greece); OTKA and NIH (Hungary); DAE and DST (India); IPM (Iran); SFI (Ireland); INFN (Italy); MSIP and NRF (Republic of Korea); LAS (Lithuania); MOE and UM (Malaysia); CINVESTAV, CONACYT, SEP, and UASLP-FAI (Mexico); MBIE (New Zealand); PAEC (Pakistan); MSHE and NSC (Poland); FCT (Portugal); JINR (Dubna); MON, RosAtom, RAS and RFBR (Russia); MESTD (Serbia); SEIDI and CPAN (Spain); Swiss Funding Agencies (Switzerland); MST (Taipei); ThEPCenter, IPST, STAR and NSTDA (Thailand); TUBITAK and TAEK (Turkey); NASU and SFFR (Ukraine); STFC (United Kingdom); DOE and NSF (USA).

Individuals have received support from the Marie-Curie program and the European Research Council and EPLANET (European Union); the Leventis Foundation; the A. P. Sloan Foundation; the Alexander von Humboldt Foundation; the Belgian Federal Science Policy Office; the Fonds pour la Formation \`a la Recherche dans l'Industrie et dans l'Agriculture (FRIA-Belgium); the Agentschap voor Innovatie door Wetenschap en Technologie (IWT-Belgium); the Ministry of Education, Youth and Sports (MEYS) of the Czech Republic; the Council of Science and Industrial Research, India; the HOMING PLUS program of Foundation for Polish Science, cofinanced from European Union, Regional Development Fund; the Compagnia di San Paolo (Torino); the Consorzio per la Fisica (Trieste); MIUR project 20108T4XTM (Italy); the Thalis and Aristeia programs cofinanced by EU-ESF and the Greek NSRF; and the National Priorities Research Program by Qatar National Research Fund.
\end{acknowledgments}

\bibliography{auto_generated}

\providecommand{\href}[2]{#2}\begingroup\raggedright\begin{thebibliography}{10}%
\makeatletter
\providecommand{\hrefCMSnoop }[0]{\@secondoftwo}%
\makeatother
\providecommand{\doi}{\texttt{doi:}\begingroup \urlstyle{tt}\Url}

\bibitem{CMS:2008zzk}
\hrefCMSnoop {}{{CMS} Collaboration, ``{The CMS experiment at the CERN LHC}'',}
  \textit{ JINST} \textbf{ 3} (2008) S08004,
\href{http://dx.doi.org/10.1088/1748-0221/3/08/S08004}{\doi{10.1088/1748-0221/3/08/S08004}}.

\bibitem{Martin:1997ns}
\hrefCMSnoop {}{S.~P. Martin, ``{A supersymmetry primer}'',} \textit{ Adv. Ser.
  Direct High Energy Phys.} \textbf{ 21} (2010) 1,
  \href{http://dx.doi.org/10.1142/9789814307505_0001}{\doi{10.1142/9789814307505_0001}},
  \href{http://www.arXiv.org/abs/hep-ph/9709356}{\texttt{arXiv:hep-ph/9709356}}.

\bibitem{edge}
I.~Hinchliffe\hrefCMSnoop {}{ {et~al.}, ``{Precision SUSY measurements at CERN
  LHC}'',} \textit{ Phys. Rev. D} \textbf{ 55} (1997) 5520,
  \href{http://dx.doi.org/10.1103/PhysRevD.55.5520}{\doi{10.1103/PhysRevD.55.5520}},
  \href{http://www.arXiv.org/abs/hep-ph/9610544}{\texttt{arXiv:hep-ph/9610544}}.

\bibitem{edge2011}
\hrefCMSnoop {}{{CMS} Collaboration, ``{Search for new physics in events with
  opposite-sign leptons, jets, and missing transverse energy in pp collisions
  at $\sqrt{s} = 7$ TeV}'',} \textit{ Phys. Lett. B} \textbf{ 718} (2013) 815,
  \href{http://dx.doi.org/10.1016/j.physletb.2012.11.036}{\doi{10.1016/j.physletb.2012.11.036}},
  \href{http://www.arXiv.org/abs/1206.3949}{\texttt{arXiv:1206.3949}}.

\bibitem{zjets2011}
\hrefCMSnoop {}{{CMS} Collaboration, ``{Search for physics beyond the standard
  model in events with a Z boson, jets, and missing transverse energy in pp
  collisions at $\sqrt{s} = 7$ TeV}'',} \textit{ Phys. Lett. B} \textbf{ 716}
  (2012) 260,
  \href{http://dx.doi.org/10.1016/j.physletb.2012.08.026}{\doi{10.1016/j.physletb.2012.08.026}},
  \href{http://www.arXiv.org/abs/1204.3774}{\texttt{arXiv:1204.3774}}.

\bibitem{ATLAS8TeVref1}
\hrefCMSnoop {}{{ATLAS} Collaboration, ``{Search for direct production of
  charginos, neutralinos and sleptons in final states with two leptons and
  missing transverse momentum in $pp$ collisions at $\sqrt{s} =$ 8 TeV with the
  ATLAS detector}'',} \textit{ JHEP} \textbf{ 05} (2014) 071,
  \href{http://dx.doi.org/10.1007/JHEP05(2014)071}{\doi{10.1007/JHEP05(2014)071}},
\href{http://www.arXiv.org/abs/1403.5294}{\texttt{arXiv:1403.5294}}.

\bibitem{ATLAS8TeVref2}
\hrefCMSnoop {}{{ATLAS} Collaboration, ``{Search for direct top-squark pair
  production in final states with two leptons in pp collisions at $\sqrt{s} =$
  8 TeV with the ATLAS detector}'',} \textit{ JHEP} \textbf{ 06} (2014) 124,
  \href{http://dx.doi.org/10.1007/JHEP06(2014)124}{\doi{10.1007/JHEP06(2014)124}},
\href{http://www.arXiv.org/abs/1403.4853}{\texttt{arXiv:1403.4853}}.

\bibitem{Alves:2011wf}
\hrefCMSnoop {}{D.~Alves {et~al.}, ``Simplified models for {LHC} new physics
  searches'',} \textit{ J. Phys. G} \textbf{ 39} (2012) 105005,
  \href{http://dx.doi.org/10.1088/0954-3899/39/10/105005}{\doi{10.1088/0954-3899/39/10/105005}},
  \href{http://www.arXiv.org/abs/1105.2838}{\texttt{arXiv:1105.2838}}.

\bibitem{Matchev:1999ft}
\hrefCMSnoop {}{K.~T. Matchev and S.~D. Thomas, ``{Higgs and Z boson signatures
  of supersymmetry}'',} \textit{ Phys. Rev. D} \textbf{ 62} (2000) 077702,
  \href{http://dx.doi.org/10.1103/PhysRevD.62.077702}{\doi{10.1103/PhysRevD.62.077702}},
\href{http://www.arXiv.org/abs/hep-ph/9908482}{\texttt{arXiv:hep-ph/9908482}}.

\bibitem{Meade:2009qv}
\hrefCMSnoop {}{P.~Meade, M.~Reece, and D.~Shih, ``{Prompt decays of general
  neutralino NLSPs at the Tevatron}'',} \textit{ JHEP} \textbf{ 05} (2010) 105,
  \href{http://dx.doi.org/10.1007/JHEP05(2010)105}{\doi{10.1007/JHEP05(2010)105}},
\href{http://www.arXiv.org/abs/0911.4130}{\texttt{arXiv:0911.4130}}.

\bibitem{ref:ewkino}
\hrefCMSnoop {}{J.~T. Ruderman and D.~Shih, ``{General neutralino NLSPs at the
  early LHC}'',} \textit{ JHEP} \textbf{ 08} (2012) 159,
  \href{http://dx.doi.org/10.1007/JHEP08(2012)159}{\doi{10.1007/JHEP08(2012)159}},
\href{http://www.arXiv.org/abs/1103.6083}{\texttt{arXiv:1103.6083}}.

\bibitem{madgraph5}
J.~Alwall\hrefCMSnoop {}{ {et~al.}, ``{MadGraph} 5: going beyond'',} \textit{
  JHEP} \textbf{ 06} (2011) 128,
  \href{http://dx.doi.org/10.1007/JHEP06(2011)128}{\doi{10.1007/JHEP06(2011)128}},
\href{http://www.arXiv.org/abs/1106.0522}{\texttt{arXiv:1106.0522}}.

\bibitem{Alwall:2014hca}
J.~Alwall\hrefCMSnoop {}{ {et~al.}, ``The automated computation of tree-level
  and next-to-leading order differential cross sections, and their matching to
  parton shower simulations'',} \textit{ JHEP} \textbf{ 07} (2014) 079,
  \href{http://dx.doi.org/10.1007/JHEP07(2014)079}{\doi{10.1007/JHEP07(2014)079}},
\href{http://www.arXiv.org/abs/1405.0301}{\texttt{arXiv:1405.0301}}.

\bibitem{Pythia}
\hrefCMSnoop {}{T.~Sj{\"o}strand, S.~Mrenna, and P.~Z. Skands, ``{PYTHIA 6.4
  physics and manual}'',} \textit{ JHEP} \textbf{ 05} (2006) 026,
  \href{http://dx.doi.org/10.1088/1126-6708/2006/05/026}{\doi{10.1088/1126-6708/2006/05/026}},
\href{http://www.arXiv.org/abs/hep-ph/0603175}{\texttt{arXiv:hep-ph/0603175}}.

\bibitem{CMSFastSim}
S.~Abdullin\hrefCMSnoop {}{ {et~al.}, ``The fast simulation of the {CMS}
  detector at {LHC}'',} in \textit{ Intl. Conf. on Computing in High Energy and
  Nuclear Physics (CHEP 2010)}.
\newblock 2011.
\newblock {J. Phys.: Conf. Ser. 331 (2012) 032049}.
\href{http://dx.doi.org/10.1088/1742-6596/331/3/032049}{\doi{10.1088/1742-6596/331/3/032049}}.

\bibitem{Geant}
\hrefCMSnoop {}{{GEANT4} Collaboration, ``{GEANT4}---a simulation toolkit'',}
  \textit{ Nucl. Instrum. Meth. A} \textbf{ 506} (2003) 250,
  \href{http://dx.doi.org/10.1016/S0168-9002(03)01368-8}{\doi{10.1016/S0168-9002(03)01368-8}}.

\bibitem{bib-nlo-nll-01}
\hrefCMSnoop {}{W.~Beenakker, R.~H{\"o}pker, M.~Spira, and P.~M. Zerwas,
  ``Squark and gluino production at hadron colliders'',} \textit{ Nucl. Phys.
  B} \textbf{ 492} (1997) 51,
  \href{http://dx.doi.org/10.1016/S0550-3213(97)80027-2}{\doi{10.1016/S0550-3213(97)80027-2}},
\href{http://www.arXiv.org/abs/hep-ph/9610490}{\texttt{arXiv:hep-ph/9610490}}.

\bibitem{bib-nlo-nll-02}
\hrefCMSnoop {}{A.~Kulesza and L.~Motyka, ``{Threshold resummation for
  squark-antisquark and gluino-pair production at the {LHC}}'',} \textit{ Phys.
  Rev. Lett.} \textbf{ 102} (2009) 111802,
  \href{http://dx.doi.org/10.1103/PhysRevLett.102.111802}{\doi{10.1103/PhysRevLett.102.111802}},
\href{http://www.arXiv.org/abs/0807.2405}{\texttt{arXiv:0807.2405}}.

\bibitem{bib-nlo-nll-03}
\hrefCMSnoop {}{A.~Kulesza and L.~Motyka, ``{Soft gluon resummation for the
  production of gluino-gluino and squark-antisquark pairs at the {LHC}}'',}
  \textit{ Phys. Rev. D} \textbf{ 80} (2009) 095004,
  \href{http://dx.doi.org/10.1103/PhysRevD.80.095004}{\doi{10.1103/PhysRevD.80.095004}},
\href{http://www.arXiv.org/abs/0905.4749}{\texttt{arXiv:0905.4749}}.

\bibitem{bib-nlo-nll-04}
W.~Beenakker\hrefCMSnoop {}{ {et~al.}, ``{Soft-gluon resummation for squark and
  gluino hadroproduction}'',} \textit{ JHEP} \textbf{ 12} (2009) 041,
  \href{http://dx.doi.org/10.1088/1126-6708/2009/12/041}{\doi{10.1088/1126-6708/2009/12/041}},
\href{http://www.arXiv.org/abs/0909.4418}{\texttt{arXiv:0909.4418}}.

\bibitem{bib-nlo-nll-05}
W.~Beenakker\hrefCMSnoop {}{ {et~al.}, ``{Squark and gluino
  hadroproduction}'',} \textit{ Int. J. Mod. Phys. A} \textbf{ 26} (2011) 2637,
  \href{http://dx.doi.org/10.1142/S0217751X11053560}{\doi{10.1142/S0217751X11053560}},
\href{http://www.arXiv.org/abs/1105.1110}{\texttt{arXiv:1105.1110}}.

\bibitem{ref:xsec}
M.~Kr{\"a}mer\hrefCMSnoop {}{ {et~al.}, ``{Supersymmetry production cross
  sections in pp collisions at $\sqrt{s}$ = 7~TeV}'',} (2012).
\href{http://www.arXiv.org/abs/1206.2892}{\texttt{arXiv:1206.2892}}.

\bibitem{EGMpaper}
\hrefCMSnoop {}{{CMS} Collaboration, ``{Performance of electron reconstruction
  and selection with the CMS detector in proton-proton collisions at sqrt(s)=8
  TeV}'',} (2015).
\href{http://www.arXiv.org/abs/1502.02701}{\texttt{arXiv:1502.02701}}.

\bibitem{MUOpaper}
\hrefCMSnoop {}{{CMS} Collaboration, ``Performance of {CMS} muon reconstruction
  in pp collision events at $\sqrt{s}=7$ {TeV}'',} \textit{ JINST} \textbf{ 7}
  (2012) P10002,
  \href{http://dx.doi.org/10.1088/1748-0221/7/10/P10002}{\doi{10.1088/1748-0221/7/10/P10002}},
\href{http://www.arXiv.org/abs/1206.4071}{\texttt{arXiv:1206.4071}}.

\bibitem{FastJet}
\hrefCMSnoop {}{M.~Cacciari, G.~P. Salam, and G.~Soyez, ``{FastJet} user
  manual'',} \textit{ Eur. Phys. J. C} \textbf{ 72} (2012) 1896,
  \href{http://dx.doi.org/10.1140/epjc/s10052-012-1896-2}{\doi{10.1140/epjc/s10052-012-1896-2}},
\href{http://www.arXiv.org/abs/1111.6097}{\texttt{arXiv:1111.6097}}.

\bibitem{Chatrchyan:2012zz}
\hrefCMSnoop {}{{CMS} Collaboration, ``{Performance of tau-lepton
  reconstruction and identification in CMS}'',} \textit{ JINST} \textbf{ 7}
  (2012) P01001,
  \href{http://dx.doi.org/10.1088/1748-0221/7/01/P01001}{\doi{10.1088/1748-0221/7/01/P01001}},
\href{http://www.arXiv.org/abs/1109.6034}{\texttt{arXiv:1109.6034}}.

\bibitem{Cacciari:2008gp}
\hrefCMSnoop {}{M.~Cacciari, G.~P. Salam, and G.~Soyez, ``{The anti-$k_t$ jet
  clustering algorithm}'',} \textit{ JHEP} \textbf{ 04} (2008) 063,
  \href{http://dx.doi.org/10.1088/1126-6708/2008/04/063}{\doi{10.1088/1126-6708/2008/04/063}},
  \href{http://www.arXiv.org/abs/0802.1189}{\texttt{arXiv:0802.1189}}.

\bibitem{Cacciari:2005hq}
\hrefCMSnoop {}{M.~Cacciari and G.~P. Salam, ``{Dispelling the N$^3$ myth for
  the $k_t$ jet-finder}'',} \textit{ Phys. Lett. B} \textbf{ 641} (2006) 57,
  \href{http://dx.doi.org/10.1016/j.physletb.2006.08.037}{\doi{10.1016/j.physletb.2006.08.037}},
  \href{http://www.arXiv.org/abs/hep-ph/0512210}{\texttt{arXiv:hep-ph/0512210}}.

\bibitem{cacciari-2008-659}
\hrefCMSnoop {}{M.~Cacciari and G.~P. Salam, ``Pileup subtraction using jet
  areas'',} \textit{ Phys. Lett. B} \textbf{ 659} (2008) 119,
  \href{http://dx.doi.org/10.1016/j.physletb.2007.09.077}{\doi{10.1016/j.physletb.2007.09.077}},
  \href{http://www.arXiv.org/abs/0707.1378}{\texttt{arXiv:0707.1378}}.

\bibitem{Li:2012wna}
\hrefCMSnoop {}{Y.~Li and F.~Petriello, ``{Combining QCD and electroweak
  corrections to dilepton production in FEWZ}'',} \textit{ Phys. Rev. D}
  \textbf{ 86} (2012) 094034,
  \href{http://dx.doi.org/10.1103/PhysRevD.86.094034}{\doi{10.1103/PhysRevD.86.094034}},
\href{http://www.arXiv.org/abs/1208.5967}{\texttt{arXiv:1208.5967}}.

\bibitem{Beringer:1900zz}
\hrefCMSnoop {}{{Particle Data Group}, K.~A. Olive {et~al.}, ``{Review of
  Particle Physics}'',} \textit{ Chin. Phys. C} \textbf{ 38} (2014) 090001,
\href{http://dx.doi.org/10.1088/1674-1137/38/9/090001}{\doi{10.1088/1674-1137/38/9/090001}}.

\bibitem{Crystal}
\href {http://www.slac.stanford.edu/pubs/slacreports/slac-r-236.html}{M.~J.
  Oreglia, ``A study of the reactions $\psi^\prime \to \gamma \gamma \psi$''}.
\newblock PhD thesis, Stanford University, 1980.
\newblock {SLAC} Report {SLAC-R-236}, see Appendix {D}.

\bibitem{James:1975dr}
\hrefCMSnoop {}{F.~James and M.~Roos, ``{Minuit: A System for Function
  Minimization and Analysis of the Parameter Errors and Correlations}'',}
  \textit{ Comput. Phys. Commun.} \textbf{ 10} (1975) 343,
\href{http://dx.doi.org/10.1016/0010-4655(75)90039-9}{\doi{10.1016/0010-4655(75)90039-9}}.

\bibitem{HiggsTool1}
\href {https://cds.cern.ch/record/1379837}{{ATLAS and CMS Collaborations},
  ``{Procedure for the LHC Higgs boson search combination in summer 2011}'',}
  Technical Report ATL-PHYS-PUB-2011-011, CMS-NOTE-2011-005, CERN, 2011.

\bibitem{Chatrchyan:2012jua}
\hrefCMSnoop {}{{CMS} Collaboration, ``{Identification of b-quark jets with the
  CMS experiment}'',} \textit{ JINST} \textbf{ 8} (2013) P04013,
  \href{http://dx.doi.org/10.1088/1748-0221/8/04/P04013}{\doi{10.1088/1748-0221/8/04/P04013}},
\href{http://www.arXiv.org/abs/1211.4462}{\texttt{arXiv:1211.4462}}.

\bibitem{CMS-PAS-LUM-13-001}
\href {http://cdsweb.cern.ch/record/1598864}{{CMS} Collaboration, ``CMS
  Luminosity Based on Pixel Cluster Counting - Summer 2013 Update'',} CMS
  Physics Analysis Summary CMS-PAS-LUM-13-001, 2013.

\bibitem{Alekhin:2011sk}
\hrefCMSnoop {}{S.~Alekhin {et~al.}, ``{The PDF4LHC Working Group Interim
  Report}'',} (2011).
\href{http://www.arXiv.org/abs/1101.0536}{\texttt{arXiv:1101.0536}}.

\bibitem{Botje:2011sn}
M.~Botje\hrefCMSnoop {}{ {et~al.}, ``{The PDF4LHC Working Group Interim
  Recommendations}'',} (2011).
\href{http://www.arXiv.org/abs/1101.0538}{\texttt{arXiv:1101.0538}}.

\bibitem{Ball:2012cx}
\hrefCMSnoop {}{{NNPDF} Collaboration, ``{Parton distributions with LHC
  data}'',} \textit{ Nucl. Phys. B} \textbf{ 867} (2013) 244,
  \href{http://dx.doi.org/10.1016/j.nuclphysb.2012.10.003}{\doi{10.1016/j.nuclphysb.2012.10.003}},
\href{http://www.arXiv.org/abs/1207.1303}{\texttt{arXiv:1207.1303}}.

\bibitem{Martin:2009iq}
\hrefCMSnoop {}{A.~D. Martin, W.~J. Stirling, R.~S. Thorne, and G.~Watt,
  ``{Parton distributions for the LHC}'',} \textit{ Eur. Phys. J. C} \textbf{
  63} (2009) 189,
  \href{http://dx.doi.org/10.1140/epjc/s10052-009-1072-5}{\doi{10.1140/epjc/s10052-009-1072-5}},
\href{http://www.arXiv.org/abs/0901.0002}{\texttt{arXiv:0901.0002}}.

\bibitem{Lai:2010vv}
H.-L. Lai\hrefCMSnoop {}{ {et~al.}, ``New parton distributions for collider
  physics'',} \textit{ Phys. Rev. D} \textbf{ 82} (2010) 074024,
  \href{http://dx.doi.org/10.1103/PhysRevD.82.074024}{\doi{10.1103/PhysRevD.82.074024}},
\href{http://www.arXiv.org/abs/1007.2241}{\texttt{arXiv:1007.2241}}.

\bibitem{1748-0221-6-11-P11002}
\hrefCMSnoop {}{{CMS} Collaboration, ``Determination of jet energy calibration
  and transverse momentum resolution in {CMS}'',} \textit{ J. Instrum.}
  \textbf{ 6} (2011) P11002,
  \href{http://dx.doi.org/10.1088/1748-0221/6/11/P11002}{\doi{10.1088/1748-0221/6/11/P11002}}.

\bibitem{Chatrchyan:2013xna}
\hrefCMSnoop {}{{CMS} Collaboration, ``{Search for top-squark pair production
  in the single-lepton final state in pp collisions at $\sqrt{s}$ = 8 TeV}'',}
  \textit{ Eur .Phys. J. C} \textbf{ 73} (2013) 2677,
  \href{http://dx.doi.org/10.1140/epjc/s10052-013-2677-2}{\doi{10.1140/epjc/s10052-013-2677-2}},
\href{http://www.arXiv.org/abs/1308.1586}{\texttt{arXiv:1308.1586}}.

\bibitem{Junk:1999kv}
\hrefCMSnoop {}{T.~Junk, ``{Confidence level computation for combining searches
  with small statistics}'',} \textit{ Nucl. Instrum. Meth. A} \textbf{ 434}
  (1999) 435,
  \href{http://dx.doi.org/10.1016/S0168-9002(99)00498-2}{\doi{10.1016/S0168-9002(99)00498-2}},
\href{http://www.arXiv.org/abs/hep-ex/9902006}{\texttt{arXiv:hep-ex/9902006}}.

\bibitem{0954-3899-28-10-313}
\hrefCMSnoop {}{A.~L. Read, ``Presentation of search results: the {$CL_s$}
  technique'',} \textit{ J. Phys. G} \textbf{ 28} (2002) 2693,
\href{http://dx.doi.org/10.1088/0954-3899/28/10/313}{\doi{10.1088/0954-3899/28/10/313}}.

\end{thebibliography}\endgroup
\cleardoublepage \appendix\section{The CMS Collaboration \label{app:collab}}\begin{sloppypar}\hyphenpenalty=5000\widowpenalty=500\clubpenalty=5000\textbf{Yerevan Physics Institute,  Yerevan,  Armenia}\\*[0pt]
V.~Khachatryan, A.M.~Sirunyan, A.~Tumasyan
\vskip\cmsinstskip
\textbf{Institut f\"{u}r Hochenergiephysik der OeAW,  Wien,  Austria}\\*[0pt]
W.~Adam, T.~Bergauer, M.~Dragicevic, J.~Er\"{o}, M.~Friedl, R.~Fr\"{u}hwirth\cmsAuthorMark{1}, V.M.~Ghete, C.~Hartl, N.~H\"{o}rmann, J.~Hrubec, M.~Jeitler\cmsAuthorMark{1}, W.~Kiesenhofer, V.~Kn\"{u}nz, M.~Krammer\cmsAuthorMark{1}, I.~Kr\"{a}tschmer, D.~Liko, I.~Mikulec, D.~Rabady\cmsAuthorMark{2}, B.~Rahbaran, H.~Rohringer, R.~Sch\"{o}fbeck, J.~Strauss, W.~Treberer-Treberspurg, W.~Waltenberger, C.-E.~Wulz\cmsAuthorMark{1}
% ...
\vskip\cmsinstskip
% ...
1:~~Also at Vienna University of Technology, Vienna, Austria\\
2:~~Also at CERN, European Organization for Nuclear Research, Geneva, Switzerland\\
% ...

\end{sloppypar}
\end{document}